\documentclass[prb,superscriptaddress,amsfonts,twocolumn,showpacs,floatfix]{revtex4}

\usepackage{amsmath}
\usepackage{amsfonts}
\usepackage{amssymb}
\usepackage{bm}
\usepackage{tabularx}
\usepackage{graphicx,color}

\def\Vec#1{\bm{#1}}
\def\Hc2{H_\mathrm{c2}}
\def\Tc{T_\mathrm{c}}

\begin{document}

\title{Thermodynamic study of gap structure and pair-breaking effect by magnetic field \\in the heavy-fermion superconductor CeCu$_2$Si$_2$}

\author{Shunichiro Kittaka}
\affiliation{Institute for Solid State Physics (ISSP), University of Tokyo, Kashiwa, Chiba 277-8581, Japan}
\author{Yuya Aoki}
\affiliation{Institute for Solid State Physics (ISSP), University of Tokyo, Kashiwa, Chiba 277-8581, Japan}
\author{Yasuyuki Shimura}
\affiliation{Institute for Solid State Physics (ISSP), University of Tokyo, Kashiwa, Chiba 277-8581, Japan}
\author{Toshiro Sakakibara}
\affiliation{Institute for Solid State Physics (ISSP), University of Tokyo, Kashiwa, Chiba 277-8581, Japan}
\author{Silvia~Seiro}
\affiliation{Max Planck Institute for Chemical Physics of Solids, 01187 Dresden, Germany}
\affiliation{Department of Chemistry and Physics of Materials, University of Salzburg, Hellbrunner Str. 34, 5020 Salzburg, Austria}
\author{Christoph~Geibel}
\affiliation{Max Planck Institute for Chemical Physics of Solids, 01187 Dresden, Germany}
\author{Frank Steglich}
\affiliation{Max Planck Institute for Chemical Physics of Solids, 01187 Dresden, Germany}
\author{Yasumasa Tsutsumi}
\affiliation{Department of Basic Science, University of Tokyo, Tokyo 153-8902, Japan}
\author{Hiroaki Ikeda}
\affiliation{Department of Physics, Ritsumeikan University, Kusatsu 525-8577, Japan}
\author{Kazushige Machida}
\affiliation{Department of Physics, Ritsumeikan University, Kusatsu 525-8577, Japan}

\date{\today}

\begin{abstract}
This paper presents the results of specific-heat and magnetization measurements, in particular their field-orientation dependence, 
on the first discovered heavy-fermion superconductor CeCu$_2$Si$_2$ ($\Tc \sim 0.6$~K).
We discuss the superconducting gap structure and 
the origin of the anomalous pair-breaking phenomena, leading e.g., to the suppression of the upper critical field $\Hc2$, found in the high-field region.
The data show that the anomalous pair breaking 
becomes prominent below about 0.15~K in any field direction, but occurs closer to $\Hc2$ for $H \parallel c$.
The presence of this anomaly is confirmed by the fact that the specific-heat and magnetization data satisfy standard thermodynamic relations. 
Concerning the gap structure, field-angle dependences of the low-temperature specific heat within the $ab$ and $ac$ planes do not show any evidence for gap nodes.
From microscopic calculations in the framework of a two-band full-gap model,
the power-law-like temperature dependences of $C$ and $1/T_1$, reminiscent of nodal superconductivity, have been reproduced reasonably.
These facts further support multiband full-gap superconductivity in CeCu$_2$Si$_2$.
\end{abstract}


\maketitle

\section{Introduction}

Pairing mechanism of heavy-fermion superconductors has aroused a considerable interest
because of a strong Coulomb repulsion between heavy quasiparticles; the conventional electron-phonon interaction is unlikely to be a pairing glue.
It has been generally considered that 
the strong Coulomb repulsion can be avoided by an unconventional pairing in which the superconducting gap has nodes. Because 
the nodal structure contains crucial information on the nature of novel pairing glues,~\cite{Sakakibara2016} determination of the superconducting gap symmetry is of primary importance. 
For instance, Ce$M$In$_5$ ($M$=Co, Rh, Ir) systems have been identified to be $d_{x^2-y^2}$-wave superconductors 
from field-angle-resolved experiments,~\cite{Izawa2001PRL,An2010PRL,Park2008PRL,Kasahara2008PRL,Kittaka2012PRB} and 
spin fluctuations associated with an antiferromagnetic quantum critical point are considered to play an important role in mediating Cooper pairs.


The first heavy-fermion superconductor CeCu$_2$Si$_2$ discovered in 1979~(Ref.~\onlinecite{Steglich1979PRL}) was also considered 
to be a nodal $d$-wave superconductor mediated by antiferromagnetic spin fluctuations.\cite{Stockert2011NatPhy} 
Indeed, the superconducting ground state (S-type) sensitively transforms to the antiferromagnetic one (A-type) 
by a slight change in the composition ratio of Cu/Si.~\cite{Seiro2010PSSB}
In the superconducting state, strong limit of the upper critical field $\Hc2$ is observed at low temperatures in any field direction.
This $\Hc2$ limit, plausibly originating from the Pauli-paramagnetic effect, and a distinct decrease of the NMR Knight shift below $\Tc$~\cite{Ueda1987JPSJ} clearly indicate formation of spin-singlet Cooper pairs.
Its superconducting gap was considered to have line nodes 
from $T^3$ dependence of the nuclear relaxation rate $1/T_1$ along with the absence of a coherence peak~\cite{Kitaoka1986JPSJ,Ishida1999PRL,Fujiwara2008JPSJ}  
and $T^2$-like temperature dependence of the specific heat~\cite{Arndt2011PRL} in the intermediate-temperature region.
These results were considered as evidence of nodal $d$-wave gap symmetry in CeCu$_2$Si$_2$, such as $d_{x^2-y^2}$ and $d_{xy}$ types.~\cite{Vieyra2011PRL,Eremin2008PRL}

However, from our recent specific-heat study,~\cite{Kittaka2014PRL} CeCu$_2$Si$_2$ has been indicated to be in a full-gap superconducting state with multiband character.
Particularly, the $H$-linear behavior in the low-temperature specific heat and its isotropic field-angle dependence under a rotating magnetic field within the $ab$ plane 
are in sharp contrast to the expected nodal $d$-wave superconductivity.
Moreover, in the high-field superconducting state, unusual features have been found in both the magnetization and specific-heat data, 
suggesting anomalous pair breaking by magnetic field.
For instance, both $\Hc2$ and diamagnetic shielding are unusually reduced on cooling and the low-temperature specific heat is oddly enhanced near $\Hc2$ (hereafter referred to as ``high-field anomaly'').

In this paper, we present the results of specific-heat $C$ and magnetization $M$ measurements 
in order to uncover the origin of the high-field anomaly and to further investigate the gap structure.
Qualitatively, both  $C(T,H)$ and $M(T,H)$ exhibit the same anomalous behavior for two field directions, $H \parallel a$ and $H \parallel c$.
More specifically, the high-field anomaly appears close to $\Hc2$ below about 0.15~K,  and is apparently more distinct in $H \parallel c$ than in $H \parallel a$. 
From the obtained data, standard thermodynamic parameters are evaluated and possible origins of the high-field anomaly are discussed.
From field-angle dependences of the low-temperature specific heat measured in a rotating magnetic field within the $ab$ and $ac$ planes, 
we establish that gap nodes are absent at least on heavy-mass bands.
We also perform microscopic calculations and clarify that a multiband full-gap model gives reasonable explanation of the power-law-like temperature variation of $C(T)$ and $T_1^{-1}(T)$ in the intermediate-temperature region, 
the behaviors which had been so far regarded as circumstantial evidence for the presence of line nodes in the gap.

\section{Experimental}

Single crystals of the S-type CeCu$_2$Si$_2$ were grown by the flux method.~\cite{Seiro2010PSSB}
A high-quality single crystal, the same sample as in Ref.~\onlinecite{Kittaka2014PRL}, was used in this study.
The sample mass is 13.4 mg weight and the dimensions are approximately 1, 2, and 1~mm along the crystalline $a$, $b$, and $c$ axes, respectively.
The dc magnetization was measured down to below 0.07~K by using a high-resolution capacitive Faraday magnetometer~\cite{Sakakibara1994JJAP} in a dilution refrigerator (Oxford Kelvinox25),
which was inserted into a superconducting magnet generating a maximum field of 150~kOe with a vertical field gradient of 500~Oe/cm.
The specific heat was measured by the standard quasi-adiabatic heat-pulse method or the relaxation method in a dilution refrigerator (Oxford Kelvinox AST Minisorb) down to 0.04~K. 
This refrigerator was inserted into the vector-magnet system that is composed of a vertical solenoid coil (up to 30~kOe), a horizontal split-pair coil (up to 50~kOe), and 
a stepper motor mounted at the top of the Dewar to rotate the refrigerator around the vertical axis.
By using this system, the field orientation was controlled three-dimensionally with high accuracy of $\sim 0.01$~deg.

\section{Results}
\subsection{Magnetization}

\begin{figure}
\includegraphics[width=3.3in]{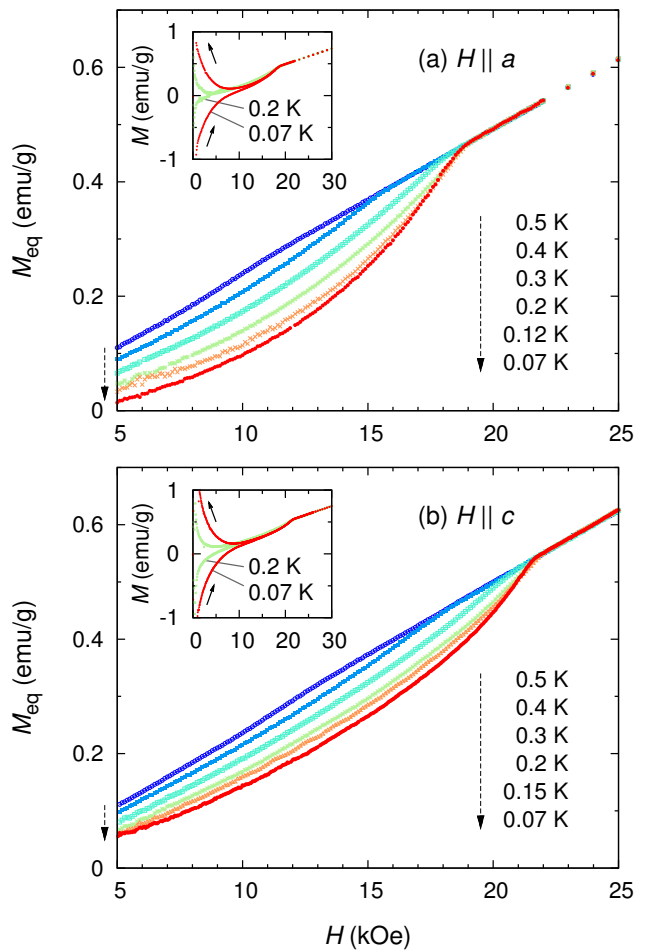}
\caption{
(Color online) Magnetic field dependence of the equilibrium magnetization $M_{\rm eq}$ at various temperatures in (a)~$H \parallel a$ and (b) $H \parallel c$.
Insets show raw magnetization curves measured at 0.07 and 0.2~K.
}
\label{MH}
\end{figure}

Raw magnetization data $M(H)$ measured at low temperatures are plotted in the insets of Figs.~\ref{MH}(a) and \ref{MH}(b) for $H \parallel a$ and $H \parallel c$, respectively.
All magnetization curves presented here were taken after zero-field cooling from a temperature well above $\Tc$.
In low fields, ordinary magnetization hysteresis due to vortex pinning is observed, 
which disturbs the detection of the lower critical field $H_{\rm c1}$ ($\sim 20$~Oe).~\cite{Rauchschwalbe1982PRL}
By contrast, in the high-field region ($H \gtrsim 10$~kOe), irreversibility of the magnetization is negligibly small,
suggesting the inclusion of defects in the present sample is small.
This hysteresis is suppressed significantly with increasing temperature.

\begin{figure}
\includegraphics[width=3.3in]{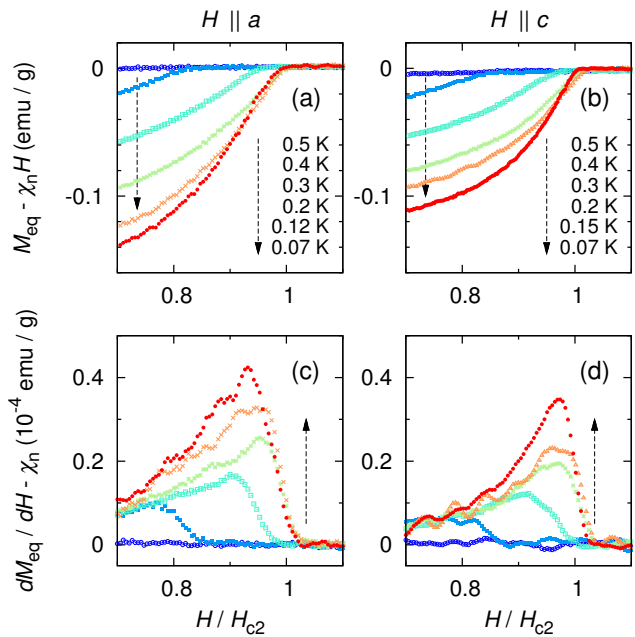}
\caption{
(Color online) Field dependence of the magnetization at several temperatures, where the paramagnetic contribution is subtracted, for (a) $H \parallel a$ and (b) $H \parallel c$.
(c), (d) Differential-susceptibility data obtained from (a) and (b), respectively.
In all the figures, the magnetic field is normalized by $\Hc2$ at 0.07~K.
}
\label{dMdH}
\end{figure}

To examine the behavior of the equilibrium magnetization, we averaged the $M(H)$ data in the increasing- and decreasing-field sequences. 
The resulting equilibrium magnetization, labelled as $M_{\rm eq}$, is plotted for several temperatures in Fig.~\ref{MH}(a) for $H \parallel a$ and Fig.~\ref{MH}(b) for $H \parallel c$.
Superconducting signature can be significantly detected in addition to the prominent paramagnetic contribution $\chi_{\rm n}H$.
In order to extract the superconducting contribution,
enlarged views of $M_{\rm eq}-\chi_{\rm n}H$ near $\Hc2$ are shown in Figs.~\ref{dMdH}(a) and \ref{dMdH}(b) for $H \parallel a$ and $H \parallel c$, respectively.
In these figures, the applied magnetic field is normalized by $\Hc2$ at $T=0.07$~K, i.e., 19~kOe for $H \parallel a$ and 21.6~kOe for $H \parallel c$.

\begin{figure}
\includegraphics[width=3.3in]{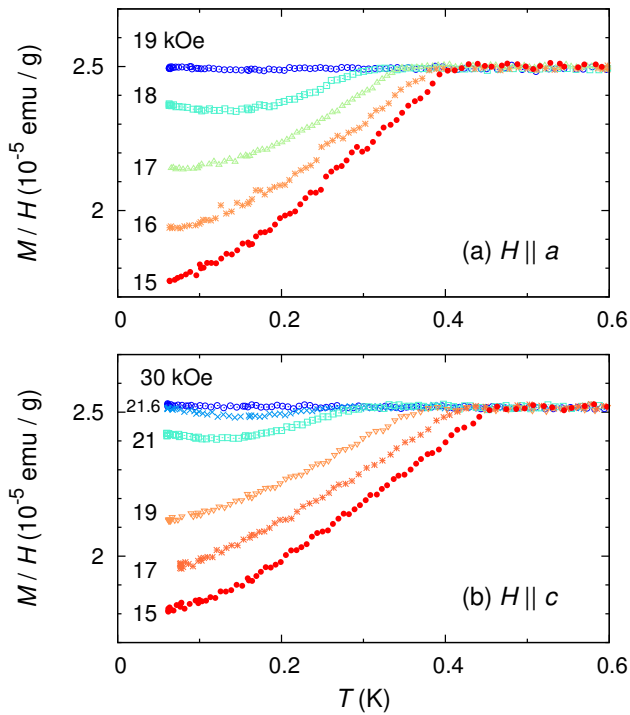}
\caption{
(Color online) Temperature dependence of $M/H$ in the zero-field-cooling process for (a) $H \parallel a$ and (b) $H \parallel c$.
}
\label{MT}
\end{figure}

With decreasing temperature, the shielding becomes pronounced below $\sim 0.95\Hc2$ accompanied by a significant increase of the slope of $M_{\rm eq}(H)$ near $\Hc2$.
The latter feature can be more clearly confirmed from development of a peak height in the field-derivative data in Figs.~\ref{dMdH}(c) and \ref{dMdH}(d).
Close inspection of the figures reveals that the onset of superconductivity,  $\Hc2$, is slightly decreasing on cooling below 0.15~K.
This tendency has already been reported from resistivity measurements.~\cite{Assmus1984PRL,Vieyra2011PRL}
Our magnetization data provide first thermodynamic evidence for this phenomenon.
In particular, the reduction of $\Hc2$ on cooling below 0.15~K is manifested in the crossing of the $M_{\rm eq}(H)$ curves for $T=0.07$ and 0.15 (0.12) K below $\Hc2$. 
This feature is more clearly seen in $H \parallel c$ than in $H \parallel a$.

Temperature dependence of the magnetization shown in Fig.~\ref{MT} also exhibits unusual behavior at low temperatures in high fields.
These data were taken after zero-field cooling, although in this field range $M(T)$ taken in  field cooling  gives almost the same result, 
as can be assured by the small hysteresis of $M(H)$ in the high-field region (see the insets of Fig.~\ref{MH}).
Under moderate magnetic fields well below $\Hc2$, e.g., at 15~kOe, $M/H$ exhibits a monotonic decease on cooling.
By contrast, in high fields slightly below $\Hc2$, $M(T)$ oddly increases at low temperatures below about 0.15~K, resulting in a distinct minimum near 0.15~K.
These results suggest that anomalous pair breaking is operative at low temperatures, particularly for $T \lesssim 0.15$~K near $\Hc2$.

\subsection{Specific heat}

\begin{figure}
\includegraphics[width=3.3in]{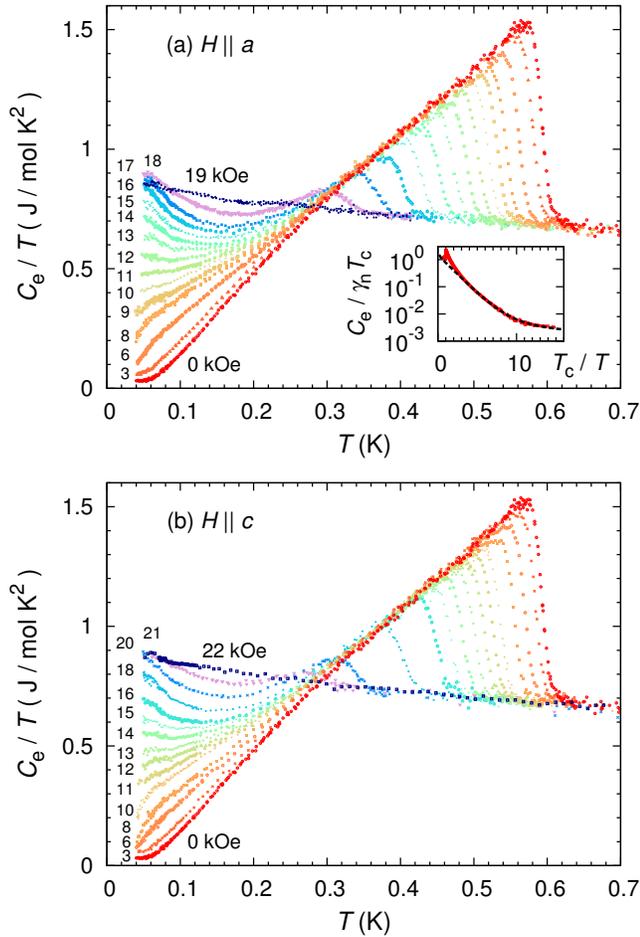}
\caption{
(Color online) Temperature dependence of $C_{\rm e}/T$ at various fields in (a) $H \parallel a $ and (b) $H \parallel c$.
Here, the nuclear specific-heat contribution is subtracted.
Inset in (a) is a semi-log plot of the zero-field $C_{\rm e}(T)$ data.
The dashed line is a fit by using the BCS function in the low-temperature regime.
}
\label{CT}
\end{figure}

Figures~\ref{CT}(a) and \ref{CT}(b) show temperature dependence of the specific-heat data, $C_{\rm e}/T$, 
measured at various magnetic fields applied parallel to the $a$ and $c$ axes, respectively. 
In all the $C_{\rm e}$ data, the nuclear contribution, $C_{\rm n}=(7.4H^2 + 0.1)/T^2$~$\mu$J/(mol K), is subtracted.~\cite{Kittaka2014PRL}
The normal-state $C_{\rm e}/T$, obtained at 19 and 22~kOe for $H\parallel a$ and $c$, respectively, shows a gradual increase on cooling,
which probably originates from three-dimensional spin-density-wave fluctuations 
in the vicinity of an antiferromagnetic quantum critical point.~\cite{Gegenwart1998,Arndt2011PRL}

In zero field, a sharp superconducting transition is observed at $\Tc=0.6$~K with a specific-heat jump of $1.25\gamma\Tc$, 
where the Sommerfeld coefficient $\gamma$ ($=0.67$~J mol$^{-1}$ K$^{-2}$) is estimated by the $C_{\rm e}/T$ value at 0.65~K.
The inset of Fig.~\ref{CT}(a) shows a semi-log plot of $C_{\rm e}/\gamma\Tc$ against $\Tc/T$.
It is clear that the low-temperature $C_{\rm e}(T)$ data ($\Tc/T\gtrsim 5$) can be well fitted by the conventional BCS expression $C(T) = A \exp(-\Delta_0/T) + \gamma_0T$ (dashed line in the same inset).
We confirmed that any power-law temperature functions, by contrast, do not adequately explain the low-temperature $C_{\rm e}(T)$ data. 
These facts imply the absence of nodes in the superconducting gap.
From the low-temperature fit, a small gap size $\Delta_0$ of 0.39~K is obtained, indicating multiband full-gap superconductivity. 
The analysis yields a small residual value of $\gamma_0=0.028$~J/(mol K$^2$),
which would be attributed to a small amount of inclusions of the non-superconducting A-type CeCu$_2$Si$_2$ in the sample.

In a weak magnetic field ($\sim 8$~kOe), $C_{\rm e}(T)/T$ exhibits a rapid 
decrease on cooling below 0.06~K for both $H \parallel a$ and $H \parallel c$.
This feature implies suppression of a minor gap. 
Above $\sim12$~kOe, on the other hand, $C_{\rm e}(T)/T$ shows an unusual increase on cooling below 0.15~K.
Very interestingly, for $H \gtrsim 0.9\Hc2$, $C_{\rm e}/T$ exceeds the normal-state value at low temperatures, 
although the system is definitely in the superconducting state as evidenced by the jump in $C_{\rm e}(T)$ near 0.3~K.
Note that this low-temperature overshoot is specific to the superconducting state and disappears above $\Hc2$. 

\begin{figure}
\includegraphics[width=3.3in]{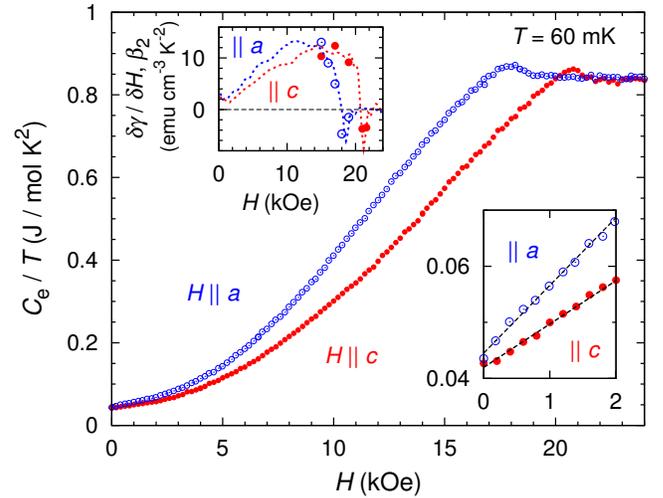}
\caption{
(Color online) 
Field dependence of $C_{\rm e}/T$ at 60~mK for $H \parallel a$ (open) and $H \parallel c$ (closed).
Lower inset shows an enlarged plot in the low-field regime.
Upper inset shows the slope of $C_{\rm e}(H)/T$ ($\delta\gamma/\delta H$; dashed lines) and the coefficient of the $T^2$ term in the $M(T,H)$ data ($\beta_2$; circles).
}
\label{CH}
\end{figure}

Figure~\ref{CH} plots field dependence of the low-temperature $C_{\rm e}/T$ measured at 0.06~K for $H \parallel a$ and $H \parallel c$.
In the low-field region, $C_{\rm e}(H)$ increases linearly with magnetic field in both field orientations, as clearly seen in the lower inset of Fig.~\ref{CH}.
In the intermediate-field region, $C_{\rm e}(H)$ exhibits upward curvature and shows a peak slightly below $\Hc2$.
These features are also confirmed in the field derivative data of $C_{\rm e}(H)/T$, labelled as $\delta \gamma / \delta H$, shown in the upper inset of Fig.~\ref{CH} by dashed lines. 
Qualitatively the same $C_{\rm e}(H)$ behavior is observed in $H \parallel a$ and $H \parallel c$.

In the low-field limit, the slope of $C_{\rm e}(H)/T$ is estimated to be 0.012 and 0.007~J/(mol K$^2$ kOe) for $H \parallel a$ and $H \parallel c$, respectively,
as represented by dashed lines in the lower inset of Fig.~\ref{CH}. 
This $H$-linear dependence is in sharp contrast to the $\sqrt{H}$ behavior predicted for nodal superconductors,~\cite{Volovik1993JETPL} 
but it is compatible with the expectation for full-gap superconductors as discussed below.

When a superconducting gap is fully open, the quasiparticle density of states increases proportionally with the number of vortex cores,
leading to $C_{\rm e} \propto H$ nearly up to $\Hc2$.
From microscopic calculations in the $T \rightarrow 0$ limit, 
$dC_{\rm e}/dH|_{H \sim 0} \sim [C_{\rm e}(\Hc2^{\rm orb})-C_{\rm e}(0)]/(0.8\Hc2^{\rm orb})$ has been predicted for an isotropic superconductor,~\cite{Nakai2002JPSJ,Nakai2004PRB}
where $\Hc2^{\rm orb}$ is an orbital-limiting field.
According to this relation, the observed initial slope implies $\Hc2^{\rm orb} \sim 84$ (140)~kOe for $H \parallel a$ ($H \parallel c$).
The result that $\Hc2^{\rm orb}\gg \Hc2$ indicates the presence of a strong Pauli paramagnetic effect as discussed in more detail in section IV-B.

One might suspect the possibility that the gradual initial slope in $C_{\rm e}(H)$ can be attributed to dirty nodal superconductivity;
in general, the initial slope of the $\sqrt{H}$ behavior in $C_{\rm e}(H)$ of nodal superconductors is easily suppressed by the thermal effect as well as the impurity-scattering effect.~\cite{Kubert1998SSC}
However, these effects would simultaneously cause significant enhancement of $\gamma_0$, which is not in the case of the present sample.

\begin{figure}
\includegraphics[width=3.3in]{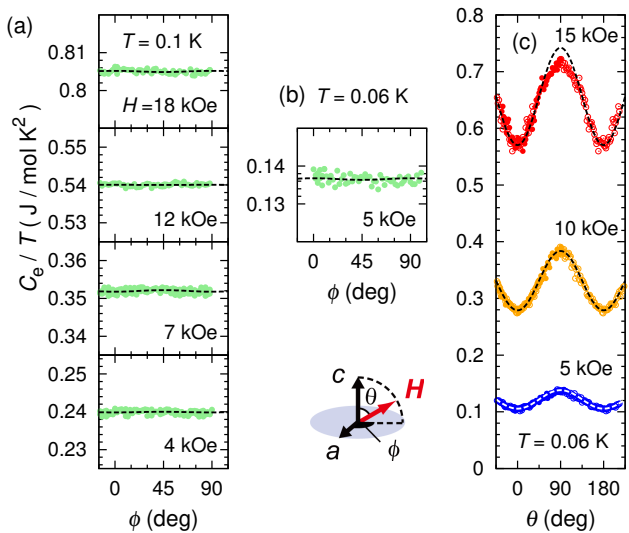}
\caption{
(Color online) 
Field angle $\phi$ dependence of $C_{\rm e}/T$ measured under a rotating magnetic field within the $ab$ plane at (a) 0.1 and (b) 0.06~K.
Here, $\phi$ is an azimuthal angle between the magnetic field and the $a$ axis. 
Dashed lines are fits to the data using $C_{\rm e}(\phi)=C_0+C_H[1-A_4\cos(4\phi)]$, 
where $C_0$ and $C_H$ are the zero-field and field-dependent parts of $C_{\rm e}$.
(c) Field-angle-resolved $C_{\rm e}/T$ at 0.06~K in a magnetic field rotated within the $ac$ plane ($\phi=90$~deg)
as a function of the field angle $\theta$ measured relative to the $c$ axis, as illustrated in the center-bottom inset.
For clarity, the data measured in the interval $-30^\circ \le \theta \le 100^\circ$ (solid symbols) are plotted repeatedly (open symbols; $\theta$ is converted to $-\theta$ and $180^\circ - \theta$).
}
\label{Cf}
\end{figure}

To obtain further evidence for nodeless superconductivity, 
we investigate the field-angle dependence of the low-temperature $C_{\rm e}$.
If nodes exist, the quasiparticle density of states, which can be detected from the low-temperature $C_{\rm e}$, should oscillate in a rotating magnetic field and 
show a local minimum when the magnetic field points along a nodal direction.~\cite{Vekhter1999PRB,Sakakibara2007JPSJ}
This is because quasiparticle excitations caused by the Doppler shift, $\delta E \propto \Vec{v}_{\rm F} \cdot \Vec{v}_{\rm s}$, are drastically suppressed at the node where $\Vec{v}_{\rm F} \perp \Vec{v}_{\rm s}$.
Here, $\Vec{v}_{\rm F}$ is the Fermi velocity and $\Vec{v}_{\rm s}$ is the supercurrent velocity circulating around vortices ($\Vec{H} \perp \Vec{v}_{\rm s}$).
Thus, the location of nodes can be detected from field-angle resolved specific-heat measurements.

We have first searched for vertical line nodes from the $\phi$-rotation experiment. 
In Figs.~\ref{Cf}(a) and \ref{Cf}(b), we plot $C_{\rm e}(\phi)/T$ measured in magnetic fields rotated within the $ab$ plane at 0.1 and 0.06~K, respectively,
where the azimuthal field angle $\phi$ is measured relative to the $a$ axis.
No oscillation that can be ascribed to anisotropic quasiparticle excitations is detected in $C_{\rm e}(\phi)$ within the resolution of the present experiment.
These results support the absence of vertical line nodes that should be expected for $d_{x^2-y^2}$- and $d_{xy}$-wave types.

Next, in order to examine the presence of horizontal line nodes, 
we investigate polar field-angle $\theta$ dependence of $C_{\rm e}/T$ in a rotating field within the $ac$ plane at 0.06~K, as shown in Fig.~\ref{Cf}(c).
Here, $\theta$ denotes the angle between the magnetic field and the $c$ axis.
Recent investigations have shown that if horizontal line nodes are present, 
$C_{\rm e}(\theta)$ is expected to exhibit a local minimum along the nodal direction and a shoulder-like or hump anomaly at certain $\theta$ tilted away from the nodal direction at low fields.~\cite{Kittaka2016JPSJ,Shimizu2016PRL,Tsutsumi2015}
Nevertheless, $C_{\rm e}(\theta)$ below 10~kOe does not show such features, as represented by dashed lines in Fig.~\ref{Cf}(c);
the data can be fitted satisfactorily by using a simple twofold function, $C(\theta,H)=A_2(H) \cos2\theta + C_0(H)$, expected to arise from the tetragonal symmetry.
Although a slight deviation from this twofold function is seen in $C_{\rm e}(\theta)$ at 15~kOe, it is attributable to the anisotropy of $\Hc2$;
the slope of $C_{\rm e}(H)$ at 15~kOe is suppressed for $H \parallel a$ whereas it is not yet for $H \parallel c$, as can be confirmed in the upper inset of Fig.~\ref{CH}.
Thus, the absence of horizontal line nodes is also well established from the $\theta$-rotation experiment.

\section{Discussion}
\subsection{Thermodynamic relations}
\begin{figure}
\includegraphics[width=3.in]{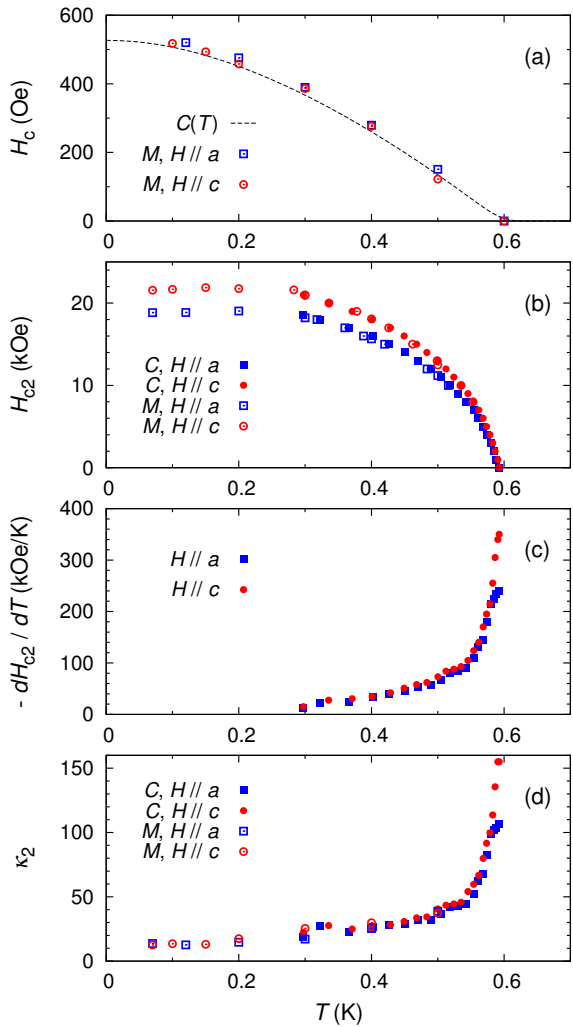}
\caption{
(Color online) 
Temperature dependences of (a) the thermodynamic critical field $H_{\rm c}$, (b) the upper critical field $\Hc2$, (c) the slope of $\Hc2(T)$, and (d) the Maki parameter $\kappa_2$.
determined from specific-heat and the magnetization measurements.
}
\label{kappa}
\end{figure}

The superconducting condensation energy $H_{\rm c}^2/8\pi$ can be evaluated from magnetization and specific-heat data by using the following relations:
\begin{align}
\frac{H_{\rm c}^2}{8\pi}&=-\int^{\Hc2}_0 [M_{\rm eq}(H)-M_{\rm n}(H)] dH \\
&=\int^{\Tc}_TdT^{\prime}\int^{\Tc}_{T^{\prime}} \frac{C_{\rm e, sc}(T^{\prime\prime})-C_{\rm e, n}(T^{\prime\prime})}{T^{\prime\prime}} dT^{\prime\prime}
\end{align}
Here, $H_{\rm c}$ is the thermodynamic critical field, $M_{\rm n}$ is the normal-state magnetization, 
and $C_{\rm e, sc}$ ($C_{\rm e, n}$) is the electronic specific heat of the zero-field superconducting state (normal state above $\Hc2$).
By assuming $M_{\rm n}(H)=\chi_{\rm n}H$, $H_{\rm c}$ is determined at each temperature from the $M_{\rm eq}(H)$ data and
its temperature dependence is plotted in Fig.~\ref{kappa}(a) by circles for $H \parallel a$ and squares for $H \parallel c$.
Likewise, $H_{\rm c}(T)$ derived from the zero-field specific-heat data is represented in Fig.~\ref{kappa}(a) by a dashed line.
These $H_{\rm c}(T)$ curves obtained from different measurements coincide well at least above 0.1~K.
This fact ensures that the present thermodynamic measurements are fully consistent with each other and
supports the absence of magnetic transitions below $\Hc2$, i.e., $M_{\rm n}(H)=\chi_{\rm n}H$.
This fact strongly suggests that the high-field anomaly has a superconducting origin.
Unfortunately, at the lowest temperature of 0.07~K, $H_{\rm c}$ could not be evaluated accurately from $M_{\rm eq}(H)$ because of the large vortex-pinning effect in the low-field region.

According to the thermodynamic Maxwell relation, the specific heat and the magnetization must satisfy
\begin{equation}
\frac{\delta}{\delta H}\frac{C(T,H)}{T}=\frac{\delta^2}{\delta T^2}M(T,H). \label{MC}
\end{equation}
Here, we examine this relation between the present thermodynamic experiments at low temperatures. 
The left side of eq.~\eqref{MC} can be replaced by the field-derivative $C_{\rm e}/T$ data at 0.06~K, 
i.e., $\delta\gamma/\delta H$ already shown in the upper inset of Fig.~\ref{CH} by dashed lines.
To evaluate the right side, temperature dependence of the magnetization [Figs.~\ref{MT}(a) and \ref{MT}(b)] is fitted by using a function $M(T,H)=M_0(H)+\frac{1}{2}\beta_2(H)T^2$ in the temperature range below 0.12~K, 
and the resulting field variation of $\beta_2$ is represented by circles in the upper inset of Fig.~\ref{CH}.
Equation~\eqref{MC} can then be reduced to  $\delta\gamma/\delta H=\beta_2$.
A good agreement between $\delta \gamma/\delta H$ and $\beta_2$ further strengthens reliability of the present thermodynamic experiment and
ensures the occurrence of the superconducting anomaly in high magnetic fields.

\subsection{Limit of the upper critical field}

The $H$-$T$ phase diagram determined from the present $C_{\rm e}(T,H)$ and $M(T,H)$ data is shown in Fig.~\ref{kappa}(b).
The results are consistent with previous studies~\cite{Assmus1984PRL,Vieyra2011PRL} and exhibit strong limit of $\Hc2$ at low temperatures in both field directions; in particular, the
strange decrease of $\Hc2$ below $\sim 0.2$~K, already reported from resistivity measurements,~\cite{Assmus1984PRL,Sheikin1998JPCM,Vieyra2011PRL} is detected by the present thermodynamic measurements as well.
It has been proposed that this phenomenon is related to the A phase; superconductivity and the A phase order compete each other.~\cite{Sheikin1998JPCM}

By using the data in Fig.~\ref{kappa}(b), we plot temperature dependence of $-d\Hc2/dT$ in Fig.~\ref{kappa}(c).
The initial slope of $\Hc2(T)$ at $\Tc$ is roughly $-240$~kOe/K for $H \parallel a$ and $-350$~kOe/K for $H \parallel c$.
On the basis of the Werthamer-Helfand-Hohenberg theory,~\cite{Helfand1966PR,Werthamer1966PR} 
the orbital-limiting field can be given by $\Hc2^{\rm orb} \sim -0.7\Tc d\Hc2/dT|_{\Tc}$. 
By using this formula, $\Hc2^{\rm orb}$ is estimated to be 100~kOe for $H \parallel a$ and 147~kOe for $H \parallel c$.
These values are in good agreements with those evaluated from the full-gap analysis of the $C_{\rm e}(H)$ data in section III-B.
These $\Hc2^{\rm orb}$ values, $\gtrsim 5$ times higher than the observed $\Hc2$, demonstrate the existence of a strong pair-breaking mechanism under magnetic fields in CeCu$_2$Si$_2$.

In order to further characterize the pair-breaking effect,
we evaluate the Ginzburg-Landau parameter $\kappa_2$ by using the following expressions: 
\begin{equation}
\frac{d(M_{\rm eq}-\chi_nH)}{dH}\biggl|_{\Hc2}=\frac{1}{4\pi\beta(2\kappa_2^2-1)}
\end{equation}
and
\begin{equation}
\frac{\Delta C}{T}\biggl|_{\Tc}=\biggl(\frac{d\Hc2}{dT}\biggl)^2\frac{1}{4\pi\beta(2\kappa_2^2-1)}.
\end{equation}
Here, $\Delta C$ is the jump of the specific heat at $\Tc(H)$ taken from Figs.~\ref{CT}(a) and \ref{CT}(b). 
The parameter $\beta$ is assumed to be 1.16 for a triangular vortex lattice.
The obtained $\kappa_2(T)$ is plotted in Fig.~\ref{kappa}(d) as a function of $T$;
$\kappa_2$ shows a rapid decrease on cooling near $\Tc$ and becomes constant at low temperatures.

A rapid decrease of $\kappa_2$ near $\Tc$ is reminiscent of the strong Pauli-paramagnetic effect.
A parameter $\alpha$ is introduced to characterize the Pauli-paramagnetic effect, and it is defined by $\alpha=\sqrt 2\Hc2^{\rm orb}(0)/H_{\rm P}(0)$, where $H_{\rm P}(0)$ denotes the Pauli limiting field at $T=0$: the upper critical field solely determined by the paramagnetic depairing effect.
For a single-band Pauli-limited superconductor with $\alpha>1$,
it is expected that $\kappa_2$ decreases rapidly near $\Tc$ and crosses zero at a finite temperature.
Below this temperature, the superconducting-normal transition at $\Hc2$ becomes first order. 
In the case of a strong Pauli-paramagnetic limit $\alpha\gg 1$, the transition at $\Hc2$ remains first order up to $\simeq 0.58\Tc$.~\cite{Sarma1963JPCS} 

The steep decrease of $\kappa_2$ below $\Tc$ in Fig.~\ref{kappa}(d) suggests $\alpha\gg 1$.
Nevertheless, $\kappa_2(T)$ levels off below 0.5~K and does not cross zero.
The absence of the first-order transition in CeCu$_2$Si$_2$ is apparently inconsistent with the conventional Pauli-limited scenario. 
We propose the unusual Pauli-paramagnetic effect to provide further support for multiband superconductivity in CeCu$_2$Si$_2$.~\cite{Kittaka2014PRL}
Indeed, the anomalous temperature dependence of $\kappa_2$ can be explained by a minimal two-band model for a Pauli-limited superconductor,~\cite{Tsutsumi2015PRB} 
whose mechanism is intuitively explained as follows. 
We define the orbital limiting field and the Pauli limiting field of band $i$ ($i=1,2$) as $\Hc2^{{\rm orb}(i)}$ and $H_{\rm P}^{(i)}$, respectively.
Assume $\Hc2^{{\rm orb}(2)}>\Hc2^{{\rm orb}(1)}$ and $H_{\rm P}^{(1)}>H_{\rm P}^{(2)}$, so that band 2 alone exhibits a first-order transition at low $T$.
The actual $\Hc2$ near $\Tc$ is governed by $\Hc2^{{\rm orb}(2)}$ and is subject to the strong paramagnetic effect, leading to a steep decrease in $\kappa_2(T)$. 
As the temperature further decreases, $\Hc2^{(2)}(T)$ of band 2 is more suppressed than $\Hc2^{(1)}(T)$ of band 1 and eventually becomes $\Hc2^{(2)}(T)<\Hc2^{(1)}(T)$.
At low temperatures, therefore, the actual $\Hc2$ is determined by $\Hc2^{(1)}(T)$ and the otherwise first-order transition of band 2 is hindered. 
Microscopic calculations based on the quasiclassical Eilenberger equations provide theoretical basis of the above argument.~\cite{Tsutsumi2015PRB}
Moreover, the calculations can also explain the low-temperature anomaly in $M(T)$ (Fig.~\ref{MT}) and $C/T$ (Figs.~\ref{CT} and \ref{CH}) at high fields as a consequence of the edge singularity of the band-2 gap.
We note that a similar field-induced anomaly has been observed in the specific heat of KFe$_2$As$_2$,~\cite{Kittaka2014JPSJ}
which also exhibits multiband superconductivity with strong Pauli-limiting behavior.~\cite{Burger2013PRB,Hardy2014JPSJ}

\subsection{Gap symmetry}

Temperature dependence of the specific heat suggests the presence of at least two fully-open gaps in CeCu$_2$Si$_2$. 
Multiband nature of CeCu$_2$Si$_2$ has also been supported by the recent results of the scanning tunneling microscope experiment.~\cite{Enayat2016PRN}
Furthermore, dependences of the low-temperature specific heat on magnetic field and its orientation have indicated the deficiency of nodal quasiparticle excitations.
On the basis of these results, it is most plausible that CeCu$_2$Si$_2$ is a full-gap superconductor.
However, the presence of nodes on light-mass bands cannot be ruled out completely only from the present study
because the specific heat is dominated by heavy-mass quasiparticles.
Therefore, identification of the gap structure by using other techniques is important, 
such as thermal-conductivity measurements and penetration depth measurements, which are powerful probes to detect light-mass quasiparticles excited from light-mass bands.

Another possibility to be considered is that the Fermi surface is absent at the locations where the gap function has nodes. 
Under such a situation, nodal quasiparticles would be totally missing.
According to the recent band-structure calculations on CeCu$_2$Si$_2$,~\cite{Kittaka2014PRL,Ikeda2015PRL} 
one electron band with the heaviest mass exists around the $X$ point, and two hole bands with relatively light mass are present around the $Z$ point.
Based on this Fermi-surface topology along with two-band full-gap nature of superconductivity, 
possibilities of $A_{2g} \:[k_xk_y(k_x^2-k_y^2)]$, $B_{1g} \:[k_x^2-k_y^2]$, $B_{2g}\: [k_{xy}]$ and $E_g$ gap functions can be safely ruled out.

If nodes are indeed absent on any bands, spin-singlet superconductivity in the tetragonal $D_{4h}$ symmetry only allows the gap symmetry of $A_{1g}$, 
i.e., a combination of the gap functions: 1, $k_x^2+k_y^2$, $k_z^2$, etc., including the case of sign-changing $A_{1g}$.
Alternatively, if we consider the possibility of orbital-selective pairing states, a fully-gapped ``$d_{x^2-y^2}+d_{xy}$'' band-mixing state might be another candidate, 
which has been recently proposed.~\cite{Pang2016}
Such gap symmetries are unprecedented in the heavy-fermion systems, and the elucidation of the pairing mechanism would contribute to deepen the understanding of unconventional superconductivity.

\vspace{0.3in}

\subsection{Multiband full-gap analyses of $C(T)$ and $T_1^{-1}(T)$}

We now turn to the temperature variation of zero-field $C_{\rm e}(T)$ and $T_1^{-1}(T)$ of CeCu$_2$Si$_2$.
The seeming power-law temperature dependences of these quantities at an intermediate-$T$ range, along with the absence of the coherence peak in $1/T_1$ just below $\Tc$, 
have been considered as solid evidence for $d$-wave superconductivity with line of nodes in this material.
A question that arises is whether the multiband full-gap model can explain these features as well. 

To answer this question, we perform microscopic calculations of the zero-field $C(T)$ and $1/T_1(T)$ by using the quasi-classical Green's functions as follows:~\cite{Nagai2008NJP,Bouquet2001EPL,Fukazawa2009JPSJ}
\begin{widetext}
\begin{equation}
\frac{(T_1T)^{-1}}{(T_1T)_{T=T_{\rm c}}^{-1}} = \frac{2}{k_{\rm B}T}\int^\infty_0 \biggl\{\biggl[\sum_i N^{\rm s}_i(E)\biggl]^2 + \biggl|\sum_i M^{\rm s}_i(E)\biggl|^2\biggl\} f(E)[1-f(E)]dE
\end{equation}
\begin{equation}
\frac{C}{\gamma_{\rm n} T}=\frac{6}{\pi^2}\frac{1}{(k_{\rm B}T)^3} \sum_i \int_0^{2\pi}\frac{d\phi}{4\pi} \int_0^\pi d\theta \sin\theta \int_0^\infty N^{\rm s}_i(E,\phi,\theta)\biggl[E^2-\frac{T}{2}\frac{d|\Delta_i(\phi,\theta)|^2}{dT}\biggl]f(E)[1-f(E)]dE.
\end{equation}
\end{widetext}
Here, the density of states is described by
\begin{align}
N^{\rm s}_i(E)&=\int_0^{2\pi}\frac{d\phi}{4\pi}\int_0^\pi d\theta\sin\theta N^{\rm s}_i(E,\phi,\theta) \\
N^{\rm s}_i(E,\phi,\theta)&=n^{\rm s}_i {\rm Re}\biggl[\frac{-i(E+i\eta)}{\sqrt{|\Delta_i(\phi,\theta)|^2-(E+i\eta)^2}}\biggl]
\end{align}
and the anomalous density of states arising from the coherence effect is written as
\begin{align}
M^{\rm s}_i(E)&=\int^{2\pi}_0\frac{d\phi}{4\pi}\int^\pi_0 d\theta\sin\theta M^{\rm s}_i(E,\phi,\theta) \\
M_i^{\rm s}(E,\phi,\theta)&=n^{\rm s}_i\Delta_i(\phi,\theta){\rm Im}\biggl[\frac{1}{\sqrt{|\Delta_i(\phi,\theta)|^2-(E+i\eta)^2}}\biggl].
\end{align}
The smearing factor $\eta$ is introduced when $1/T_1$ is calculated ($\eta=0$ when $C$ is calculated).
The parameter $n_i$ denotes the weight of the $i$-th band.
In the present analyses, a two-band full-gap model is used for simplicity; 
two superconducting gaps $\Delta_i(\phi,\theta)$ ($i=1$ and 2) are assumed to be independent of $\phi$ and $\theta$.
Temperature dependence of $\Delta_i$ is obtained by solving the gap equation at each temperature. 

\begin{figure}
\includegraphics[width=3.3in]{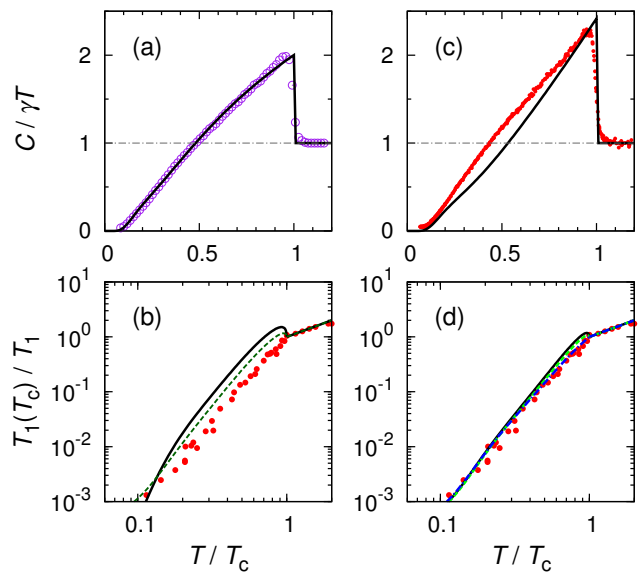}
\caption{
(Color online) 
Two-band full-gap analyses of (a) $C(T)$ and (b) $1/T_1(T)$ by using model A.
Analyses of (c) $C(T)$ and (d) $1/T_1(T)$ by using models B and C.
The calculated results are represented by lines.
Solid circles in (c) are experimental data of $C_{\rm e}/\gamma T$. 
Open circles in (a) represent modified specific-heat data, $[C_{\rm e}(T, 0~{\rm T})-C_{\rm e}(T,3~{\rm T})]/\gamma^\ast T+1$, which was introduced in Ref.~\onlinecite{Kittaka2014PRL}.
Circles in (b) and (d) are experimental data of $1/T_1$ taken from the previous report~\cite{Ishida1999PRL} on polycrystalline CeCu$_{2.05}$Si$_2$.
Solid and dashed lines in (b) represent the calculated results with $\eta/k_{\rm B}\Tc=0.01$ and 0.1, respectively.
Solid and dashed lines in (d) are the calculated results by models B and C with $\eta/k_{\rm B}\Tc=0.1$, respectively.
Models B and C provide the same results of $C(T)$.
A dot-dashed line in (d) is the calculated result by model C with temperature-dependent smearing factor, $\eta(T)/k_{\rm B}\Tc=0.1+0.5(T/\Tc)^2$.
}
\label{calc}
\end{figure}

First, we focus on model~A in Table~I which was proposed in the previous report.~\cite{Kittaka2014PRL} 
As shown in Fig.~\ref{calc}(a), $C/\gamma T$ of model~A (solid line) can reproduce sufficiently the modified zero-field specific-heat data, 
$[C_{\rm e}(T, 0~{\rm T})-C_{\rm e}(T,3~{\rm T})]/\gamma^\ast T+1$ (open circles), which satisfy entropy balance in the BCS framework;
the temperature-dependent part in the normal-state $C_{\rm e}/T$ is embedded in the superconducting $C_{\rm e}/T$ by subtracting the data at 3~T for $H \parallel a$.~\cite{Kittaka2014PRL} 
Here, $\gamma^\ast$ is equal to $C_{\rm e}(3~{\rm T})-C_{\rm e}(0~{\rm T})$ at $T \rightarrow 0$ ($=0.84$ J mol$^{-1}$ K$^{-2}$). 
This result demonstrates that the two-band full-gap model is not inconsequent for CeCu$_2$Si$_2$.

\begin{table}[b]
\begin{center}
\renewcommand{\arraystretch}{1.4}
\begin{tabular*}{6cm}{@{\extracolsep{\fill}}ccc}\hline\hline
Model & $\Delta_1/k_{\rm B}\Tc$ & $\Delta_2/k_{\rm B}\Tc$ \\ \hline
A & 1.76 & 0.7 \\
B & 2.1 & 0.8 \\
C & 2.1 & -0.8 \\
\hline\hline
\end{tabular*}
\caption{
Superconducting gaps $\Delta_i$ ($i=1$ and 2) used in the present two-band full-gap analyses.
In all the models, weights of primary ($\Delta_1$) and secondary ($\Delta_2$) bands are 65\% and 35\% of the total density of states, respectively, i.e., $n_1=0.65$ and $n_2=0.35$.
}
\label{para}
\end{center}
\end{table}

Next, we calculate $1/T_1(T)$ by using the same parameters of model~A.
Calculated results of $1/T_1(T)$ for $\eta/k_{\rm B}\Tc=0.01$ and 0.1 are represented in Fig.~\ref{calc}(b) by solid and dashed lines, respectively. 
Experimental data of $1/T_1(T)$, shown in the same figure by circles, are taken from the previous report~\cite{Ishida1999PRL} on polycrystalline S-type CeCu$_{2.05}$Si$_2$.
In the high-temperature region below $\Tc$, a clear enhancement of $1/T_1$, i.e., a coherence peak, 
appears clearly in the calculated result for $\eta/k_{\rm B}\Tc=0.01$. 
Although this enhancement is suppressed by increasing $\eta$, 
the calculated $1/T_1$ for $\eta/k_{\rm B}\Tc=0.1$ is still larger than the experimental result in wide temperature range.

In order to reproduce the strong reduction of $1/T_1$ just below $\Tc$, both $\Delta_1$ and $\Delta_2$ are increased in model~B.
Then, the calculated result [solid line in Fig.~\ref{calc}(d)] agrees approximately with the experimental result.
Here, we adopt $\eta/k_{\rm B}\Tc=0.1$.
This increase of $\Delta_i$ leads to a larger specific-heat jump at $\Tc$.
In Fig.~\ref{calc}(c), the calculated $C/\gamma T$ by model B is shown by a solid line.
The specific-heat jump, $\Delta C/\gamma T$, calculated by model B is 1.4 whereas $\Delta C/\gamma T=1.0$ by model A.
However, $\Delta C/\gamma T=1.4$ matches better with the jump in the experimental $C_{\rm e}/\gamma T$ shown in Fig.~\ref{calc}(c) by circles;
a jump in the modified specific-heat data [open circles in Fig.~\ref{calc}(a)] is underestimated because of $\gamma^\ast > \gamma$.
Thus, the gap sizes in model B are quantitatively reasonable to reproduce $C_{\rm e}(T)$ as well.

Finally, in order to examine a possibility of $s_\pm$ superconductivity, we here adopt model C in which signs of two gaps are opposite.
Calculated result of $1/T_1(T)$ by model C is shown in Fig.~\ref{calc}(d) by a dashed line; 
the coherence peak is slightly suppressed compared with that of the $s_{++}$ model (solid line, model B).
Nevertheless, difference between them is not so significant.
This is because the term $M_1^{\rm s}(E)M_2^{\rm s}(E)$, which makes the difference between $s_\pm$ and $s_{++}$, is not large for the present parameters.
Note that $C(T)$ calculated by model~C is the same as that by model B because the gap amplitude, which is the same for both cases, governs the calculated $C(T)$.
These results suggest that both $s_\pm$ and $s_{++}$ models are allowed for CeCu$_2$Si$_2$.

The lack of the coherence peak in the experimental $1/T_1$, which cannot be fully reproduced even by models B and C, 
might be due to some damping of quasiparticles.
Indeed, strong quasiparticle damping is possible in CeCu$_2$Si$_2$ particularly near $\Tc$ 
because inelastic spin fluctuations, a good candidate of the pairing interaction, are enhanced by temperature roughly proportional to $T^2$.~\cite{Stockert2011NatPhy}
If we assume that $\eta$ also increases proportionally to $T^2$ up to $\Tc$, reflecting the temperature-dependent pairing interaction,
a coherence peak almost diminishes, as demonstrated in Fig.~\ref{calc}(d) by a dot-dashed line [model C with $\eta(T)/k_{\rm B}\Tc=0.1+0.5(T/\Tc)^2$].
Thus, the behavior of $1/T_1(T)$, reminiscent of nodal superconductivity, can be explained reasonably in the framework of a multiband full-gap model.

\section{Summary}
We have performed low-temperature magnetization and specific-heat measurements on a high-quality single crystal of the S-type CeCu$_2$Si$_2$.
The high-field anomaly in the superconducting state has been detected from both thermodynamic measurements that satisfy standard thermodynamic relations.
It has been revealed that this anomaly becomes prominent below roughly 0.15~K and occurs more abrupt near $\Hc2$ in $H \parallel c$.
From azimuthal and polar field-angle dependences of the low-temperature specific heat, 
the absence of vertical and horizontal line nodes in the superconducting gap on heavy-mass bands has been established.
We have also demonstrated that a two-band full-gap model can reproduce temperature dependences of the zero-field $C_{\rm e}(T)$ and $1/T_1(T)$ in the wide temperature range,
which had been considered to be hallmarks of nodal superconductivity.
These results  help establish multiband full-gap superconductivity of CeCu$_2$Si$_2$.

\acknowledgments
We thank K. Ishida for providing us experimental data of $1/T_1$ and giving fruitful discussion.
This work was supported by a Grant-in-Aid for Scientific Research on Innovative Areas ``J-Physics'' (15H05883) from MEXT,
and KAKENHI (15K05158, 15H03682, 26400360, 15H05745, 15K17715, 15J05698) from JSPS.

\bibliographystyle{apsrev3}
\bibliography{C:/usr/local/share/texmf/bibref/ref_CeCu2Si2_ver2.0.bib}

\begin{thebibliography}{44}
\expandafter\ifx\csname natexlab\endcsname\relax\def\natexlab#1{#1}\fi
\expandafter\ifx\csname bibnamefont\endcsname\relax
  \def\bibnamefont#1{#1}\fi
\expandafter\ifx\csname bibfnamefont\endcsname\relax
  \def\bibfnamefont#1{#1}\fi
\expandafter\ifx\csname citenamefont\endcsname\relax
  \def\citenamefont#1{#1}\fi
\expandafter\ifx\csname url\endcsname\relax
  \def\url#1{\texttt{#1}}\fi
\expandafter\ifx\csname urlprefix\endcsname\relax\def\urlprefix{URL }\fi
\providecommand{\bibinfo}[2]{#2}
\providecommand{\eprint}[2][]{\url{#2}}

\bibitem[{\citenamefont{Sakakibara et~al.}()\citenamefont{Sakakibara, Kittaka,
  and Machida}}]{Sakakibara2016}
\bibinfo{author}{\bibfnamefont{T.}~\bibnamefont{Sakakibara}},
  \bibinfo{author}{\bibfnamefont{S.}~\bibnamefont{Kittaka}}, \bibnamefont{and}
  \bibinfo{author}{\bibfnamefont{K.}~\bibnamefont{Machida}}, \bibinfo{journal}{Rep. Prog. Phys.} \textbf{\bibinfo{volume}{79}},
  \bibinfo{pages}{094002} (\bibinfo{year}{2016}).

\bibitem[{\citenamefont{Izawa et~al.}(2001)\citenamefont{Izawa, Yamaguchi,
  Matsuda, Shishido, Settai, and Onuki}}]{Izawa2001PRL}
\bibinfo{author}{\bibfnamefont{K.}~\bibnamefont{Izawa}},
  \bibinfo{author}{\bibfnamefont{H.}~\bibnamefont{Yamaguchi}},
  \bibinfo{author}{\bibfnamefont{Y.}~\bibnamefont{Matsuda}},
  \bibinfo{author}{\bibfnamefont{H.}~\bibnamefont{Shishido}},
  \bibinfo{author}{\bibfnamefont{R.}~\bibnamefont{Settai}}, \bibnamefont{and}
  \bibinfo{author}{\bibfnamefont{Y.}~\bibnamefont{Onuki}},
  \bibinfo{journal}{Phys. Rev. Lett.} \textbf{\bibinfo{volume}{87}},
  \bibinfo{pages}{057002} (\bibinfo{year}{2001}).

\bibitem[{\citenamefont{An et~al.}(2010)\citenamefont{An, Sakakibara, Settai,
  Onuki, Hiragi, Ichioka, and Machida}}]{An2010PRL}
\bibinfo{author}{\bibfnamefont{K.}~\bibnamefont{An}},
  \bibinfo{author}{\bibfnamefont{T.}~\bibnamefont{Sakakibara}},
  \bibinfo{author}{\bibfnamefont{R.}~\bibnamefont{Settai}},
  \bibinfo{author}{\bibfnamefont{Y.}~\bibnamefont{Onuki}},
  \bibinfo{author}{\bibfnamefont{M.}~\bibnamefont{Hiragi}},
  \bibinfo{author}{\bibfnamefont{M.}~\bibnamefont{Ichioka}}, \bibnamefont{and}
  \bibinfo{author}{\bibfnamefont{K.}~\bibnamefont{Machida}},
  \bibinfo{journal}{Phys. Rev. Lett.} \textbf{\bibinfo{volume}{104}},
  \bibinfo{pages}{037002} (\bibinfo{year}{2010}).

\bibitem[{\citenamefont{Park et~al.}(2008)\citenamefont{Park, Bauer, and
  Thompson}}]{Park2008PRL}
\bibinfo{author}{\bibfnamefont{T.}~\bibnamefont{Park}},
  \bibinfo{author}{\bibfnamefont{E.~D.} \bibnamefont{Bauer}}, \bibnamefont{and}
  \bibinfo{author}{\bibfnamefont{J.~D.} \bibnamefont{Thompson}},
  \bibinfo{journal}{Phys. Rev. Lett.} \textbf{\bibinfo{volume}{101}},
  \bibinfo{pages}{177002} (\bibinfo{year}{2008}).

\bibitem[{\citenamefont{Kasahara et~al.}(2008)\citenamefont{Kasahara, Iwasawa,
  Shimizu, Shishido, Shibauchi, Vekhter, and Matsuda}}]{Kasahara2008PRL}
\bibinfo{author}{\bibfnamefont{Y.}~\bibnamefont{Kasahara}},
  \bibinfo{author}{\bibfnamefont{T.}~\bibnamefont{Iwasawa}},
  \bibinfo{author}{\bibfnamefont{Y.}~\bibnamefont{Shimizu}},
  \bibinfo{author}{\bibfnamefont{H.}~\bibnamefont{Shishido}},
  \bibinfo{author}{\bibfnamefont{T.}~\bibnamefont{Shibauchi}},
  \bibinfo{author}{\bibfnamefont{I.}~\bibnamefont{Vekhter}}, \bibnamefont{and}
  \bibinfo{author}{\bibfnamefont{Y.}~\bibnamefont{Matsuda}},
  \bibinfo{journal}{Phys. Rev. Lett.} \textbf{\bibinfo{volume}{100}},
  \bibinfo{pages}{207003} (\bibinfo{year}{2008}).

\bibitem[{\citenamefont{Kittaka et~al.}(2012)\citenamefont{Kittaka, Aoki,
  Sakakibara, Sakai, Nakatsuji, Tsutsumi, Ichioka, and
  Machida}}]{Kittaka2012PRB}
\bibinfo{author}{\bibfnamefont{S.}~\bibnamefont{Kittaka}},
  \bibinfo{author}{\bibfnamefont{Y.}~\bibnamefont{Aoki}},
  \bibinfo{author}{\bibfnamefont{T.}~\bibnamefont{Sakakibara}},
  \bibinfo{author}{\bibfnamefont{A.}~\bibnamefont{Sakai}},
  \bibinfo{author}{\bibfnamefont{S.}~\bibnamefont{Nakatsuji}},
  \bibinfo{author}{\bibfnamefont{Y.}~\bibnamefont{Tsutsumi}},
  \bibinfo{author}{\bibfnamefont{M.}~\bibnamefont{Ichioka}}, \bibnamefont{and}
  \bibinfo{author}{\bibfnamefont{K.}~\bibnamefont{Machida}},
  \bibinfo{journal}{Phys. Rev. B} \textbf{\bibinfo{volume}{85}},
  \bibinfo{pages}{060505} (\bibinfo{year}{2012}).

\bibitem[{\citenamefont{Steglich et~al.}(1979)\citenamefont{Steglich, Aarts,
  Bredl, Lieke, Meschede, Franz, and Sch\"afer}}]{Steglich1979PRL}
\bibinfo{author}{\bibfnamefont{F.}~\bibnamefont{Steglich}},
  \bibinfo{author}{\bibfnamefont{J.}~\bibnamefont{Aarts}},
  \bibinfo{author}{\bibfnamefont{C.~D.} \bibnamefont{Bredl}},
  \bibinfo{author}{\bibfnamefont{W.}~\bibnamefont{Lieke}},
  \bibinfo{author}{\bibfnamefont{D.}~\bibnamefont{Meschede}},
  \bibinfo{author}{\bibfnamefont{W.}~\bibnamefont{Franz}}, \bibnamefont{and}
  \bibinfo{author}{\bibfnamefont{H.}~\bibnamefont{Sch\"afer}},
  \bibinfo{journal}{Phys. Rev. Lett.} \textbf{\bibinfo{volume}{43}},
  \bibinfo{pages}{1892} (\bibinfo{year}{1979}).

\bibitem[{\citenamefont{Stockert et~al.}(2011)\citenamefont{Stockert, Arndt,
  Faulhaber, Geibel, Jeevan, Kirchner, Loewenhaupt, Schmalzl, Schmidt, Si
  et~al.}}]{Stockert2011NatPhy}
\bibinfo{author}{\bibfnamefont{O.}~\bibnamefont{Stockert}},
  \bibinfo{author}{\bibfnamefont{J.}~\bibnamefont{Arndt}},
  \bibinfo{author}{\bibfnamefont{E.}~\bibnamefont{Faulhaber}},
  \bibinfo{author}{\bibfnamefont{C.}~\bibnamefont{Geibel}},
  \bibinfo{author}{\bibfnamefont{H.~S.} \bibnamefont{Jeevan}},
  \bibinfo{author}{\bibfnamefont{S.}~\bibnamefont{Kirchner}},
  \bibinfo{author}{\bibfnamefont{M.}~\bibnamefont{Loewenhaupt}},
  \bibinfo{author}{\bibfnamefont{K.}~\bibnamefont{Schmalzl}},
  \bibinfo{author}{\bibfnamefont{W.}~\bibnamefont{Schmidt}},
  \bibinfo{author}{\bibfnamefont{Q.}~\bibnamefont{Si}}, \bibnamefont{and}
  \bibinfo{author}{\bibfnamefont{F.}~\bibnamefont{Steglich}},
  \bibinfo{journal}{Nat. Phys.} \textbf{\bibinfo{volume}{7}},
  \bibinfo{pages}{119} (\bibinfo{year}{2011}).

\bibitem[{\citenamefont{Seiro et~al.}(2010)\citenamefont{Seiro, Deppe, Jeevan,
  Burkhardt, and Geibel}}]{Seiro2010PSSB}
\bibinfo{author}{\bibfnamefont{S.}~\bibnamefont{Seiro}},
  \bibinfo{author}{\bibfnamefont{M.}~\bibnamefont{Deppe}},
  \bibinfo{author}{\bibfnamefont{H.}~\bibnamefont{Jeevan}},
  \bibinfo{author}{\bibfnamefont{U.}~\bibnamefont{Burkhardt}},
  \bibnamefont{and} \bibinfo{author}{\bibfnamefont{C.}~\bibnamefont{Geibel}},
  \bibinfo{journal}{Phys. Status Solidi B \label{Seiro2010PSSB}}
  \textbf{\bibinfo{volume}{247}}, \bibinfo{pages}{614} (\bibinfo{year}{2010}).

\bibitem[{\citenamefont{Ueda et~al.}(1987)\citenamefont{Ueda, Kitaoka, Yamada,
  Kohori, Kohara, and Asayama}}]{Ueda1987JPSJ}
\bibinfo{author}{\bibfnamefont{K.}~\bibnamefont{Ueda}},
  \bibinfo{author}{\bibfnamefont{Y.}~\bibnamefont{Kitaoka}},
  \bibinfo{author}{\bibfnamefont{H.}~\bibnamefont{Yamada}},
  \bibinfo{author}{\bibfnamefont{Y.}~\bibnamefont{Kohori}},
  \bibinfo{author}{\bibfnamefont{T.}~\bibnamefont{Kohara}}, \bibnamefont{and}
  \bibinfo{author}{\bibfnamefont{K.}~\bibnamefont{Asayama}},
  \bibinfo{journal}{J. Phys. Soc. Jpn.} \textbf{\bibinfo{volume}{56}},
  \bibinfo{pages}{867} (\bibinfo{year}{1987}).

\bibitem[{\citenamefont{Kitaoka et~al.}(1986)\citenamefont{Kitaoka, Ueda,
  Fujiwara, Arimoto, Iida, and Asayama}}]{Kitaoka1986JPSJ}
\bibinfo{author}{\bibfnamefont{Y.}~\bibnamefont{Kitaoka}},
  \bibinfo{author}{\bibfnamefont{K.}~\bibnamefont{Ueda}},
  \bibinfo{author}{\bibfnamefont{K.}~\bibnamefont{Fujiwara}},
  \bibinfo{author}{\bibfnamefont{H.}~\bibnamefont{Arimoto}},
  \bibinfo{author}{\bibfnamefont{H.}~\bibnamefont{Iida}}, \bibnamefont{and}
  \bibinfo{author}{\bibfnamefont{K.}~\bibnamefont{Asayama}},
  \bibinfo{journal}{J. Phys. Soc. Jpn.} \textbf{\bibinfo{volume}{55}},
  \bibinfo{pages}{723} (\bibinfo{year}{1986}).

\bibitem[{\citenamefont{Ishida et~al.}(1999)\citenamefont{Ishida, Kawasaki,
  Tabuchi, Kashima, Kitaoka, Asayama, Geibel, and Steglich}}]{Ishida1999PRL}
\bibinfo{author}{\bibfnamefont{K.}~\bibnamefont{Ishida}},
  \bibinfo{author}{\bibfnamefont{Y.}~\bibnamefont{Kawasaki}},
  \bibinfo{author}{\bibfnamefont{K.}~\bibnamefont{Tabuchi}},
  \bibinfo{author}{\bibfnamefont{K.}~\bibnamefont{Kashima}},
  \bibinfo{author}{\bibfnamefont{Y.}~\bibnamefont{Kitaoka}},
  \bibinfo{author}{\bibfnamefont{K.}~\bibnamefont{Asayama}},
  \bibinfo{author}{\bibfnamefont{C.}~\bibnamefont{Geibel}}, \bibnamefont{and}
  \bibinfo{author}{\bibfnamefont{F.}~\bibnamefont{Steglich}},
  \bibinfo{journal}{Phys. Rev. Lett.} \textbf{\bibinfo{volume}{82}},
  \bibinfo{pages}{5353} (\bibinfo{year}{1999}).

\bibitem[{\citenamefont{Fujiwara et~al.}(2008)\citenamefont{Fujiwara, Hata,
  Kobayashi, Miyoshi, Takeuchi, Shimaoka, Kotegawa, Kobayashi, Geibel, and
  Steglich}}]{Fujiwara2008JPSJ}
\bibinfo{author}{\bibfnamefont{K.}~\bibnamefont{Fujiwara}},
  \bibinfo{author}{\bibfnamefont{Y.}~\bibnamefont{Hata}},
  \bibinfo{author}{\bibfnamefont{K.}~\bibnamefont{Kobayashi}},
  \bibinfo{author}{\bibfnamefont{K.}~\bibnamefont{Miyoshi}},
  \bibinfo{author}{\bibfnamefont{J.}~\bibnamefont{Takeuchi}},
  \bibinfo{author}{\bibfnamefont{Y.}~\bibnamefont{Shimaoka}},
  \bibinfo{author}{\bibfnamefont{H.}~\bibnamefont{Kotegawa}},
  \bibinfo{author}{\bibfnamefont{T.~C.} \bibnamefont{Kobayashi}},
  \bibinfo{author}{\bibfnamefont{C.}~\bibnamefont{Geibel}}, \bibnamefont{and}
  \bibinfo{author}{\bibfnamefont{F.}~\bibnamefont{Steglich}},
  \bibinfo{journal}{J. Phys. Soc. Jpn.} \textbf{\bibinfo{volume}{77}},
  \bibinfo{pages}{123711} (\bibinfo{year}{2008}).

\bibitem[{\citenamefont{Arndt et~al.}(2011)\citenamefont{Arndt, Stockert,
  Schmalzl, Faulhaber, Jeevan, Geibel, Schmidt, Loewenhaupt, and
  Steglich}}]{Arndt2011PRL}
\bibinfo{author}{\bibfnamefont{J.}~\bibnamefont{Arndt}},
  \bibinfo{author}{\bibfnamefont{O.}~\bibnamefont{Stockert}},
  \bibinfo{author}{\bibfnamefont{K.}~\bibnamefont{Schmalzl}},
  \bibinfo{author}{\bibfnamefont{E.}~\bibnamefont{Faulhaber}},
  \bibinfo{author}{\bibfnamefont{H.~S.} \bibnamefont{Jeevan}},
  \bibinfo{author}{\bibfnamefont{C.}~\bibnamefont{Geibel}},
  \bibinfo{author}{\bibfnamefont{W.}~\bibnamefont{Schmidt}},
  \bibinfo{author}{\bibfnamefont{M.}~\bibnamefont{Loewenhaupt}},
  \bibnamefont{and} \bibinfo{author}{\bibfnamefont{F.}~\bibnamefont{Steglich}},
  \bibinfo{journal}{Phys. Rev. Lett.} \textbf{\bibinfo{volume}{106}},
  \bibinfo{pages}{246401} (\bibinfo{year}{2011}).

\bibitem[{\citenamefont{Vieyra et~al.}(2011)\citenamefont{Vieyra, Oeschler,
  Seiro, Jeevan, Geibel, Parker, and Steglich}}]{Vieyra2011PRL}
\bibinfo{author}{\bibfnamefont{H.~A.} \bibnamefont{Vieyra}},
  \bibinfo{author}{\bibfnamefont{N.}~\bibnamefont{Oeschler}},
  \bibinfo{author}{\bibfnamefont{S.}~\bibnamefont{Seiro}},
  \bibinfo{author}{\bibfnamefont{H.~S.} \bibnamefont{Jeevan}},
  \bibinfo{author}{\bibfnamefont{C.}~\bibnamefont{Geibel}},
  \bibinfo{author}{\bibfnamefont{D.}~\bibnamefont{Parker}}, \bibnamefont{and}
  \bibinfo{author}{\bibfnamefont{F.}~\bibnamefont{Steglich}},
  \bibinfo{journal}{Phys. Rev. Lett.\label{Vieyra2011PRL}}
  \textbf{\bibinfo{volume}{106}}, \bibinfo{pages}{207001}
  (\bibinfo{year}{2011}).

\bibitem[{\citenamefont{Eremin et~al.}(2008)\citenamefont{Eremin, Zwicknagl,
  Thalmeier, and Fulde}}]{Eremin2008PRL}
\bibinfo{author}{\bibfnamefont{I.}~\bibnamefont{Eremin}},
  \bibinfo{author}{\bibfnamefont{G.}~\bibnamefont{Zwicknagl}},
  \bibinfo{author}{\bibfnamefont{P.}~\bibnamefont{Thalmeier}},
  \bibnamefont{and} \bibinfo{author}{\bibfnamefont{P.}~\bibnamefont{Fulde}},
  \bibinfo{journal}{Phys. Rev. Lett.} \textbf{\bibinfo{volume}{101}},
  \bibinfo{pages}{187001} (\bibinfo{year}{2008}).

\bibitem[{\citenamefont{Kittaka
  et~al.}(2014{\natexlab{a}})\citenamefont{Kittaka, Aoki, Shimura, Sakakibara,
  Seiro, Geibel, Steglich, Ikeda, and Machida}}]{Kittaka2014PRL}
\bibinfo{author}{\bibfnamefont{S.}~\bibnamefont{Kittaka}},
  \bibinfo{author}{\bibfnamefont{Y.}~\bibnamefont{Aoki}},
  \bibinfo{author}{\bibfnamefont{Y.}~\bibnamefont{Shimura}},
  \bibinfo{author}{\bibfnamefont{T.}~\bibnamefont{Sakakibara}},
  \bibinfo{author}{\bibfnamefont{S.}~\bibnamefont{Seiro}},
  \bibinfo{author}{\bibfnamefont{C.}~\bibnamefont{Geibel}},
  \bibinfo{author}{\bibfnamefont{F.}~\bibnamefont{Steglich}},
  \bibinfo{author}{\bibfnamefont{H.}~\bibnamefont{Ikeda}}, \bibnamefont{and}
  \bibinfo{author}{\bibfnamefont{K.}~\bibnamefont{Machida}},
  \bibinfo{journal}{Phys. Rev. Lett.} \textbf{\bibinfo{volume}{112}},
  \bibinfo{pages}{067002} (\bibinfo{year}{2014}{\natexlab{a}}).

\bibitem[{\citenamefont{Sakakibara et~al.}(1994)\citenamefont{Sakakibara,
  Mitamura, Tayama, and Amitsuka}}]{Sakakibara1994JJAP}
\bibinfo{author}{\bibfnamefont{T.}~\bibnamefont{Sakakibara}},
  \bibinfo{author}{\bibfnamefont{H.}~\bibnamefont{Mitamura}},
  \bibinfo{author}{\bibfnamefont{T.}~\bibnamefont{Tayama}}, \bibnamefont{and}
  \bibinfo{author}{\bibfnamefont{H.}~\bibnamefont{Amitsuka}},
  \bibinfo{journal}{Jpn. J. Appl. Phys.} \textbf{\bibinfo{volume}{33}},
  \bibinfo{pages}{5067} (\bibinfo{year}{1994}).

\bibitem[{\citenamefont{Rauchschwalbe et~al.}(1982)\citenamefont{Rauchschwalbe,
  Lieke, Bredl, Steglich, Aarts, Martini, and Mota}}]{Rauchschwalbe1982PRL}
\bibinfo{author}{\bibfnamefont{U.}~\bibnamefont{Rauchschwalbe}},
  \bibinfo{author}{\bibfnamefont{W.}~\bibnamefont{Lieke}},
  \bibinfo{author}{\bibfnamefont{C.~D.} \bibnamefont{Bredl}},
  \bibinfo{author}{\bibfnamefont{F.}~\bibnamefont{Steglich}},
  \bibinfo{author}{\bibfnamefont{J.}~\bibnamefont{Aarts}},
  \bibinfo{author}{\bibfnamefont{K.~M.} \bibnamefont{Martini}},
  \bibnamefont{and} \bibinfo{author}{\bibfnamefont{A.~C.} \bibnamefont{Mota}},
  \bibinfo{journal}{Phys. Rev. Lett.} \textbf{\bibinfo{volume}{49}},
  \bibinfo{pages}{1448} (\bibinfo{year}{1982}).

\bibitem[{\citenamefont{Assmus et~al.}(1984)\citenamefont{Assmus, Herrmann,
  Rauchschwalbe, Riegel, Lieke, Spille, Horn, Weber, Steglich, and
  Cordier}}]{Assmus1984PRL}
\bibinfo{author}{\bibfnamefont{W.}~\bibnamefont{Assmus}},
  \bibinfo{author}{\bibfnamefont{M.}~\bibnamefont{Herrmann}},
  \bibinfo{author}{\bibfnamefont{U.}~\bibnamefont{Rauchschwalbe}},
  \bibinfo{author}{\bibfnamefont{S.}~\bibnamefont{Riegel}},
  \bibinfo{author}{\bibfnamefont{W.}~\bibnamefont{Lieke}},
  \bibinfo{author}{\bibfnamefont{H.}~\bibnamefont{Spille}},
  \bibinfo{author}{\bibfnamefont{S.}~\bibnamefont{Horn}},
  \bibinfo{author}{\bibfnamefont{G.}~\bibnamefont{Weber}},
  \bibinfo{author}{\bibfnamefont{F.}~\bibnamefont{Steglich}}, \bibnamefont{and}
  \bibinfo{author}{\bibfnamefont{G.}~\bibnamefont{Cordier}},
  \bibinfo{journal}{Phys. Rev. Lett.} \textbf{\bibinfo{volume}{52}},
  \bibinfo{pages}{469} (\bibinfo{year}{1984}).

\bibitem[{\citenamefont{Gegenwart et~al.}(1998)\citenamefont{Gegenwart,
  Langhammer, Geibel, Helfrich, Lang, Sparn, Steglich, Horn, Donnevert, Link
  et~al.}}]{Gegenwart1998}
\bibinfo{author}{\bibfnamefont{P.}~\bibnamefont{Gegenwart}},
  \bibinfo{author}{\bibfnamefont{C.}~\bibnamefont{Langhammer}},
  \bibinfo{author}{\bibfnamefont{C.}~\bibnamefont{Geibel}},
  \bibinfo{author}{\bibfnamefont{R.}~\bibnamefont{Helfrich}},
  \bibinfo{author}{\bibfnamefont{M.}~\bibnamefont{Lang}},
  \bibinfo{author}{\bibfnamefont{G.}~\bibnamefont{Sparn}},
  \bibinfo{author}{\bibfnamefont{F.}~\bibnamefont{Steglich}},
  \bibinfo{author}{\bibfnamefont{R.}~\bibnamefont{Horn}},
  \bibinfo{author}{\bibfnamefont{L.}~\bibnamefont{Donnevert}},
  \bibinfo{author}{\bibfnamefont{A.}~\bibnamefont{Link}}, \bibnamefont{and}
  \bibinfo{author}{\bibfnamefont{W.}~\bibnamefont{Assmus}},
  \bibinfo{journal}{Phys. Rev. Lett.} \textbf{\bibinfo{volume}{81}},
  \bibinfo{pages}{1501} (\bibinfo{year}{1998}).

\bibitem[{\citenamefont{Volovik}(1993)}]{Volovik1993JETPL}
\bibinfo{author}{\bibfnamefont{G.~E.} \bibnamefont{Volovik}},
  \bibinfo{journal}{JETP Lett.} \textbf{\bibinfo{volume}{58}},
  \bibinfo{pages}{469} (\bibinfo{year}{1993}).

\bibitem[{\citenamefont{Nakai et~al.}(2002)\citenamefont{Nakai, Ichioka, and
  Machida}}]{Nakai2002JPSJ}
\bibinfo{author}{\bibfnamefont{N.}~\bibnamefont{Nakai}},
  \bibinfo{author}{\bibfnamefont{M.}~\bibnamefont{Ichioka}}, \bibnamefont{and}
  \bibinfo{author}{\bibfnamefont{K.}~\bibnamefont{Machida}},
  \bibinfo{journal}{J. Phys. Soc. Jpn.} \textbf{\bibinfo{volume}{71}},
  \bibinfo{pages}{23} (\bibinfo{year}{2002}).

\bibitem[{\citenamefont{Nakai et~al.}(2004)\citenamefont{Nakai, Miranovi\'{c},
  Ichioka, and Machida}}]{Nakai2004PRB}
\bibinfo{author}{\bibfnamefont{N.}~\bibnamefont{Nakai}},
  \bibinfo{author}{\bibfnamefont{P.}~\bibnamefont{Miranovi\'{c}}},
  \bibinfo{author}{\bibfnamefont{M.}~\bibnamefont{Ichioka}}, \bibnamefont{and}
  \bibinfo{author}{\bibfnamefont{K.}~\bibnamefont{Machida}},
  \bibinfo{journal}{Phys. Rev. B} \textbf{\bibinfo{volume}{70}},
  \bibinfo{pages}{100503} (\bibinfo{year}{2004}).

\bibitem[{\citenamefont{K\"{u}bert and Hirschfeld}(1998)}]{Kubert1998SSC}
\bibinfo{author}{\bibfnamefont{C.}~\bibnamefont{K\"{u}bert}} \bibnamefont{and}
  \bibinfo{author}{\bibfnamefont{P.~J.} \bibnamefont{Hirschfeld}},
  \bibinfo{journal}{Solid State Commun.} \textbf{\bibinfo{volume}{105}},
  \bibinfo{pages}{459} (\bibinfo{year}{1998}).

\bibitem[{\citenamefont{Vekhter et~al.}(1999)\citenamefont{Vekhter, Hirschfeld,
  Carbotte, and Nicol}}]{Vekhter1999PRB}
\bibinfo{author}{\bibfnamefont{I.}~\bibnamefont{Vekhter}},
  \bibinfo{author}{\bibfnamefont{P.~J.} \bibnamefont{Hirschfeld}},
  \bibinfo{author}{\bibfnamefont{J.~P.} \bibnamefont{Carbotte}},
  \bibnamefont{and} \bibinfo{author}{\bibfnamefont{E.~J.} \bibnamefont{Nicol}},
  \bibinfo{journal}{Phys. Rev. B} \textbf{\bibinfo{volume}{59}},
  \bibinfo{pages}{R9023} (\bibinfo{year}{1999}).

\bibitem[{\citenamefont{Sakakibara et~al.}(2007)\citenamefont{Sakakibara,
  Yamada, Custers, Yano, Tayama, Aoki, and Machida}}]{Sakakibara2007JPSJ}
\bibinfo{author}{\bibfnamefont{T.}~\bibnamefont{Sakakibara}},
  \bibinfo{author}{\bibfnamefont{A.}~\bibnamefont{Yamada}},
  \bibinfo{author}{\bibfnamefont{J.}~\bibnamefont{Custers}},
  \bibinfo{author}{\bibfnamefont{K.}~\bibnamefont{Yano}},
  \bibinfo{author}{\bibfnamefont{T.}~\bibnamefont{Tayama}},
  \bibinfo{author}{\bibfnamefont{H.}~\bibnamefont{Aoki}}, \bibnamefont{and}
  \bibinfo{author}{\bibfnamefont{K.}~\bibnamefont{Machida}},
  \bibinfo{journal}{J. Phys. Soc. Jpn.} \textbf{\bibinfo{volume}{76}},
  \bibinfo{pages}{051004} (\bibinfo{year}{2007}).

\bibitem[{\citenamefont{Kittaka et~al.}(2016)\citenamefont{Kittaka, Shimizu,
  Sakakibara, Haga, Yamamoto, {$\bar{\mathrm{O}}$}nuki, Tsutsumi, Nomoto,
  Ikeda, and Machida}}]{Kittaka2016JPSJ}
\bibinfo{author}{\bibfnamefont{S.}~\bibnamefont{Kittaka}},
  \bibinfo{author}{\bibfnamefont{Y.}~\bibnamefont{Shimizu}},
  \bibinfo{author}{\bibfnamefont{T.}~\bibnamefont{Sakakibara}},
  \bibinfo{author}{\bibfnamefont{Y.}~\bibnamefont{Haga}},
  \bibinfo{author}{\bibfnamefont{E.}~\bibnamefont{Yamamoto}},
  \bibinfo{author}{\bibfnamefont{Y.}~\bibnamefont{{$\bar{\mathrm{O}}$}nuki}},
  \bibinfo{author}{\bibfnamefont{Y.}~\bibnamefont{Tsutsumi}},
  \bibinfo{author}{\bibfnamefont{T.}~\bibnamefont{Nomoto}},
  \bibinfo{author}{\bibfnamefont{H.}~\bibnamefont{Ikeda}}, \bibnamefont{and}
  \bibinfo{author}{\bibfnamefont{K.}~\bibnamefont{Machida}},
  \bibinfo{journal}{J. Phys. Soc. Jpn.} \textbf{\bibinfo{volume}{85}},
  \bibinfo{pages}{033704} (\bibinfo{year}{2016}).

\bibitem[{\citenamefont{Shimizu et~al.}(2016)\citenamefont{Shimizu, Kittaka,
  Sakakibara, Tsutsumi, Nomoto, Ikeda, Machida, Homma, and
  Aoki}}]{Shimizu2016PRL}
\bibinfo{author}{\bibfnamefont{Y.}~\bibnamefont{Shimizu}},
  \bibinfo{author}{\bibfnamefont{S.}~\bibnamefont{Kittaka}},
  \bibinfo{author}{\bibfnamefont{T.}~\bibnamefont{Sakakibara}},
  \bibinfo{author}{\bibfnamefont{Y.}~\bibnamefont{Tsutsumi}},
  \bibinfo{author}{\bibfnamefont{T.}~\bibnamefont{Nomoto}},
  \bibinfo{author}{\bibfnamefont{H.}~\bibnamefont{Ikeda}},
  \bibinfo{author}{\bibfnamefont{K.}~\bibnamefont{Machida}},
  \bibinfo{author}{\bibfnamefont{Y.}~\bibnamefont{Homma}}, \bibnamefont{and}
  \bibinfo{author}{\bibfnamefont{D.}~\bibnamefont{Aoki}},
  \bibinfo{journal}{Phys. Rev. Lett.} \textbf{\bibinfo{volume}{117}},
  \bibinfo{pages}{037001} (\bibinfo{year}{2016}).

\bibitem[{Tsu()}]{Tsutsumi2015}
\bibinfo{note}{Y. Tsutsumi, T. Nomoto, H. Ikeda, and K. Machida,
  arXiv:1604.02806.}

\bibitem[{\citenamefont{Sheikin et~al.}(1998)\citenamefont{Sheikin,
  Braithwaite, Brison, Buzdin, and Assmus}}]{Sheikin1998JPCM}
\bibinfo{author}{\bibfnamefont{I.}~\bibnamefont{Sheikin}},
  \bibinfo{author}{\bibfnamefont{D.}~\bibnamefont{Braithwaite}},
  \bibinfo{author}{\bibfnamefont{J.-P.} \bibnamefont{Brison}},
  \bibinfo{author}{\bibfnamefont{A.~I.} \bibnamefont{Buzdin}},
  \bibnamefont{and} \bibinfo{author}{\bibfnamefont{W.}~\bibnamefont{Assmus}},
  \bibinfo{journal}{J. Phys.: Condens. Matter} \textbf{\bibinfo{volume}{10}},
  \bibinfo{pages}{L749} (\bibinfo{year}{1998}).

\bibitem[{\citenamefont{Helfand and Werthamer}(1966)}]{Helfand1966PR}
\bibinfo{author}{\bibfnamefont{E.}~\bibnamefont{Helfand}} \bibnamefont{and}
  \bibinfo{author}{\bibfnamefont{N.~R.} \bibnamefont{Werthamer}},
  \bibinfo{journal}{Phys. Rev.} \textbf{\bibinfo{volume}{147}},
  \bibinfo{pages}{288} (\bibinfo{year}{1966}).

\bibitem[{\citenamefont{Werthamer et~al.}(1966)\citenamefont{Werthamer,
  Helfand, and Hohenberg}}]{Werthamer1966PR}
\bibinfo{author}{\bibfnamefont{N.~R.} \bibnamefont{Werthamer}},
  \bibinfo{author}{\bibfnamefont{E.}~\bibnamefont{Helfand}}, \bibnamefont{and}
  \bibinfo{author}{\bibfnamefont{P.~C.} \bibnamefont{Hohenberg}},
  \bibinfo{journal}{Phys. Rev.} \textbf{\bibinfo{volume}{147}},
  \bibinfo{pages}{295} (\bibinfo{year}{1966}).

\bibitem[{\citenamefont{Sarma}(1963)}]{Sarma1963JPCS}
\bibinfo{author}{\bibfnamefont{G.}~\bibnamefont{Sarma}}, \bibinfo{journal}{J.
  Phys. Chem. Solids} \textbf{\bibinfo{volume}{24}}, \bibinfo{pages}{1029}
  (\bibinfo{year}{1963}).

\bibitem[{\citenamefont{Tsutsumi et~al.}(2015)\citenamefont{Tsutsumi, Machida,
  and Ichioka}}]{Tsutsumi2015PRB}
\bibinfo{author}{\bibfnamefont{Y.}~\bibnamefont{Tsutsumi}},
  \bibinfo{author}{\bibfnamefont{K.}~\bibnamefont{Machida}}, \bibnamefont{and}
  \bibinfo{author}{\bibfnamefont{M.}~\bibnamefont{Ichioka}},
  \bibinfo{journal}{Phys. Rev. B} \textbf{\bibinfo{volume}{92}},
  \bibinfo{pages}{020502} (\bibinfo{year}{2015}).

\bibitem[{\citenamefont{Kittaka
  et~al.}(2014{\natexlab{b}})\citenamefont{Kittaka, Aoki, Kase, Sakakibara,
  Saito, Fukazawa, Kohori, Kihou, Lee, Iyo et~al.}}]{Kittaka2014JPSJ}
\bibinfo{author}{\bibfnamefont{S.}~\bibnamefont{Kittaka}},
  \bibinfo{author}{\bibfnamefont{Y.}~\bibnamefont{Aoki}},
  \bibinfo{author}{\bibfnamefont{N.}~\bibnamefont{Kase}},
  \bibinfo{author}{\bibfnamefont{T.}~\bibnamefont{Sakakibara}},
  \bibinfo{author}{\bibfnamefont{T.}~\bibnamefont{Saito}},
  \bibinfo{author}{\bibfnamefont{H.}~\bibnamefont{Fukazawa}},
  \bibinfo{author}{\bibfnamefont{Y.}~\bibnamefont{Kohori}},
  \bibinfo{author}{\bibfnamefont{K.}~\bibnamefont{Kihou}},
  \bibinfo{author}{\bibfnamefont{C.~H.} \bibnamefont{Lee}},
  \bibinfo{author}{\bibfnamefont{A.}~\bibnamefont{Iyo}},
  \bibinfo{author}{\bibfnamefont{H.}~\bibnamefont{Eisaki}},
  \bibinfo{author}{\bibfnamefont{K.}~\bibnamefont{Deguchi}},
  \bibinfo{author}{\bibfnamefont{N.~K.} \bibnamefont{Sato}},
  \bibinfo{author}{\bibfnamefont{Y.}~\bibnamefont{Tsutsumi}}, \bibnamefont{and}
  \bibinfo{author}{\bibfnamefont{K.}~\bibnamefont{Machida}},
  \bibinfo{journal}{J. Phys. Soc. Jpn.} \textbf{\bibinfo{volume}{83}},
  \bibinfo{pages}{013704} (\bibinfo{year}{2014}{\natexlab{b}}).

\bibitem[{\citenamefont{Burger et~al.}(2013)\citenamefont{Burger, Hardy, Aoki,
  B\"ohmer, Eder, Heid, Wolf, Schweiss, Fromknecht, Jackson
  et~al.}}]{Burger2013PRB}
\bibinfo{author}{\bibfnamefont{P.}~\bibnamefont{Burger}},
  \bibinfo{author}{\bibfnamefont{F.}~\bibnamefont{Hardy}},
  \bibinfo{author}{\bibfnamefont{D.}~\bibnamefont{Aoki}},
  \bibinfo{author}{\bibfnamefont{A.~E.} \bibnamefont{B\"ohmer}},
  \bibinfo{author}{\bibfnamefont{R.}~\bibnamefont{Eder}},
  \bibinfo{author}{\bibfnamefont{R.}~\bibnamefont{Heid}},
  \bibinfo{author}{\bibfnamefont{T.}~\bibnamefont{Wolf}},
  \bibinfo{author}{\bibfnamefont{P.}~\bibnamefont{Schweiss}},
  \bibinfo{author}{\bibfnamefont{R.}~\bibnamefont{Fromknecht}},
  \bibinfo{author}{\bibfnamefont{M.~J.} \bibnamefont{Jackson}},
  \bibinfo{author}{\bibfnamefont{C.}~\bibnamefont{Paulsen}}, \bibnamefont{and}
  \bibinfo{author}{\bibfnamefont{C.}~\bibnamefont{Meingast}},
  \bibinfo{journal}{Phys. Rev. B} \textbf{\bibinfo{volume}{88}},
  \bibinfo{pages}{014517} (\bibinfo{year}{2013}).

\bibitem[{\citenamefont{Hardy et~al.}(2014)\citenamefont{Hardy, Eder, Jackson,
  Aoki, Paulsen, Wolf, Burger, B\"ohmer, Schweiss, Adelmann
  et~al.}}]{Hardy2014JPSJ}
\bibinfo{author}{\bibfnamefont{F.}~\bibnamefont{Hardy}},
  \bibinfo{author}{\bibfnamefont{R.}~\bibnamefont{Eder}},
  \bibinfo{author}{\bibfnamefont{M.}~\bibnamefont{Jackson}},
  \bibinfo{author}{\bibfnamefont{D.}~\bibnamefont{Aoki}},
  \bibinfo{author}{\bibfnamefont{C.}~\bibnamefont{Paulsen}},
  \bibinfo{author}{\bibfnamefont{T.}~\bibnamefont{Wolf}},
  \bibinfo{author}{\bibfnamefont{P.}~\bibnamefont{Burger}},
  \bibinfo{author}{\bibfnamefont{A.}~\bibnamefont{B\"ohmer}},
  \bibinfo{author}{\bibfnamefont{P.}~\bibnamefont{Schweiss}},
  \bibinfo{author}{\bibfnamefont{P.}~\bibnamefont{Adelmann}},
  \bibinfo{author}{\bibfnamefont{R.~A.} \bibnamefont{Fisher}},
  \bibnamefont{and} \bibinfo{author}{\bibfnamefont{C.}~\bibnamefont{Meingast}},
  \bibinfo{journal}{J. Phys. Soc. Jpn.} \textbf{\bibinfo{volume}{83}},
  \bibinfo{pages}{014711} (\bibinfo{year}{2014}).

\bibitem[{\citenamefont{Enayat et~al.}(2016)\citenamefont{Enayat, Sun,
  Maldonado, Suderow, Seiro, Geibel, Wirth, Steglich, and
  Wahl}}]{Enayat2016PRN}
\bibinfo{author}{\bibfnamefont{M.}~\bibnamefont{Enayat}},
  \bibinfo{author}{\bibfnamefont{Z.}~\bibnamefont{Sun}},
  \bibinfo{author}{\bibfnamefont{A.}~\bibnamefont{Maldonado}},
  \bibinfo{author}{\bibfnamefont{H.}~\bibnamefont{Suderow}},
  \bibinfo{author}{\bibfnamefont{S.}~\bibnamefont{Seiro}},
  \bibinfo{author}{\bibfnamefont{C.}~\bibnamefont{Geibel}},
  \bibinfo{author}{\bibfnamefont{S.}~\bibnamefont{Wirth}},
  \bibinfo{author}{\bibfnamefont{F.}~\bibnamefont{Steglich}}, \bibnamefont{and}
  \bibinfo{author}{\bibfnamefont{P.}~\bibnamefont{Wahl}},
  \bibinfo{journal}{Phys. Rev. B} \textbf{\bibinfo{volume}{93}},
  \bibinfo{pages}{045123} (\bibinfo{year}{2016}).

\bibitem[{\citenamefont{Ikeda et~al.}(2015)\citenamefont{Ikeda, Suzuki, and
  Arita}}]{Ikeda2015PRL}
\bibinfo{author}{\bibfnamefont{H.}~\bibnamefont{Ikeda}},
  \bibinfo{author}{\bibfnamefont{M.-T.} \bibnamefont{Suzuki}},
  \bibnamefont{and} \bibinfo{author}{\bibfnamefont{R.}~\bibnamefont{Arita}},
  \bibinfo{journal}{Phys. Rev. Lett.} \textbf{\bibinfo{volume}{114}},
  \bibinfo{pages}{147003} (\bibinfo{year}{2015}).

\bibitem[{Pan()}]{Pang2016}
\bibinfo{note}{G. M. Pang, M. Smidman, J. L. Zhang, L. Jiao, Z. F. Weng, E. M.
  Nica, Y. Chen, W. B. Jiang, Y. J. Zhang, H. S. Jeevan, P. Gegenwart, F.
  Steglich, Q. Si, H. Q. Yuan, arXiv:1605.04786.}

\bibitem[{\citenamefont{Nagai et~al.}(2008)\citenamefont{Nagai, Hayashi, Nakai,
  Nakamura, Okumura, and Machida}}]{Nagai2008NJP}
\bibinfo{author}{\bibfnamefont{Y.}~\bibnamefont{Nagai}},
  \bibinfo{author}{\bibfnamefont{N.}~\bibnamefont{Hayashi}},
  \bibinfo{author}{\bibfnamefont{N.}~\bibnamefont{Nakai}},
  \bibinfo{author}{\bibfnamefont{H.}~\bibnamefont{Nakamura}},
  \bibinfo{author}{\bibfnamefont{M.}~\bibnamefont{Okumura}}, \bibnamefont{and}
  \bibinfo{author}{\bibfnamefont{M.}~\bibnamefont{Machida}},
  \bibinfo{journal}{New J. Phys.} \textbf{\bibinfo{volume}{10}},
  \bibinfo{pages}{103026} (\bibinfo{year}{2008}).

\bibitem[{\citenamefont{Bouquet et~al.}(2001)\citenamefont{Bouquet, Wang,
  Fisher, Hinks, Jorgensen, Junod, and Phillips}}]{Bouquet2001EPL}
\bibinfo{author}{\bibfnamefont{F.}~\bibnamefont{Bouquet}},
  \bibinfo{author}{\bibfnamefont{Y.}~\bibnamefont{Wang}},
  \bibinfo{author}{\bibfnamefont{R.~A.} \bibnamefont{Fisher}},
  \bibinfo{author}{\bibfnamefont{D.~G.} \bibnamefont{Hinks}},
  \bibinfo{author}{\bibfnamefont{J.~D.} \bibnamefont{Jorgensen}},
  \bibinfo{author}{\bibfnamefont{A.}~\bibnamefont{Junod}}, \bibnamefont{and}
  \bibinfo{author}{\bibfnamefont{N.~E.} \bibnamefont{Phillips}},
  \bibinfo{journal}{EuroPhys. Lett.} \textbf{\bibinfo{volume}{56}},
  \bibinfo{pages}{856} (\bibinfo{year}{2001}).

\bibitem[{\citenamefont{Fukazawa et~al.}(2009)\citenamefont{Fukazawa, Yamada,
  Kondo, Saito, Kohori, Kuga, Matsumoto, Nakatsuji, Kito, Shirage
  et~al.}}]{Fukazawa2009JPSJ}
\bibinfo{author}{\bibfnamefont{H.}~\bibnamefont{Fukazawa}},
  \bibinfo{author}{\bibfnamefont{Y.}~\bibnamefont{Yamada}},
  \bibinfo{author}{\bibfnamefont{K.}~\bibnamefont{Kondo}},
  \bibinfo{author}{\bibfnamefont{T.}~\bibnamefont{Saito}},
  \bibinfo{author}{\bibfnamefont{Y.}~\bibnamefont{Kohori}},
  \bibinfo{author}{\bibfnamefont{K.}~\bibnamefont{Kuga}},
  \bibinfo{author}{\bibfnamefont{Y.}~\bibnamefont{Matsumoto}},
  \bibinfo{author}{\bibfnamefont{S.}~\bibnamefont{Nakatsuji}},
  \bibinfo{author}{\bibfnamefont{H.}~\bibnamefont{Kito}},
  \bibinfo{author}{\bibfnamefont{P.~M.} \bibnamefont{Shirage}},
  \bibinfo{author}{\bibfnamefont{K.}~\bibnamefont{Kihou}},
  \bibinfo{author}{\bibfnamefont{N.}~\bibnamefont{Takeshita}},
  \bibinfo{author}{\bibfnamefont{C.-H.} \bibnamefont{Lee}},
  \bibinfo{author}{\bibfnamefont{A.}~\bibnamefont{Iyo}}, \bibnamefont{and}
  \bibinfo{author}{\bibfnamefont{H.}~\bibnamefont{Eisaki}},
  \bibinfo{journal}{J. Phys. Soc. Jpn.} \textbf{\bibinfo{volume}{78}},
  \bibinfo{pages}{083712} (\bibinfo{year}{2009}).

\end{thebibliography}

\end{document}